\documentclass[a4paper,12pt]{article}
\usepackage{epsfig}
\usepackage{color}
\usepackage[numbers,sort&compress]{natbib}

\newlength{\dinwidth}
\newlength{\dinmargin}
\begin{document}
\title{  Studying of $B_s^0-\bar{B}_s^0$ mixing and $B_{s}\to
K^{(*)-}K^{(*)+}$ decays within supersymmetry }
\author{Ru-Min Wang$^{1}$\thanks{E-mail: ruminwang@gmail.com},~~
Yuan-Guo Xu$^{1}$\thanks{E-mail: yuanguox@gmail.com}, ~~Qin
Chang$^{2}$\thanks{E-Mail: changqin@htu.cn},~~Ya-Dong
Yang$^{3,4}$\thanks{E-Mail: yangyd@iopp.ccnu.edu.cn}
 \\
{\scriptsize {$^1$ \it College of Physics and Electronic
Engineering, Xinyang Normal University,
 Xinyang, Henan 464000, China}}
 \\
 {\scriptsize  {$^2$ \it Department of Physics, Henan Normal University, Xinxiang, Henan 453007, P. R. China}}
\\
 {\scriptsize  {$^3$ \it Institute of Particle Physics, Huazhong Normal University, Wuhan,
Hubei 430079, P. R. China }}
\\
 {\scriptsize  {$^4$ \it Key Laboratory of Quark $\&$ Lepton Physics,
Ministry of Education, P.R. China }}}

 \maketitle\vspace{-1cm}
\begin{abstract}

Recent results from CDF and D{\O} collaborations favor a large CP
asymmetry in $B_s^0-\bar{B}_s^0$ mixing, while the standard model
prediction is very small. Such a large phase may imply sizable new
physics effects in $B_s^0-\bar{B}_s^0$ mixing. We compute the
gluino-mediated supersymmetry contributions to $B_s^0-\bar{B}_s^0$
mixing, $B_{s}\to K^{(*)-}K^{(*)+}$ and $B\to X_s \gamma$ decays in
the frame of the mass insertion approximation. Combining the
constraints of $\Delta M_s,~\Delta\Gamma_s,~\phi^{J/\psi\phi}_s$,
$\mathcal{B}(B_{s}\to K^{-}K^{+})$ and $\mathcal{B}(B\to X_s
\gamma)$, we find that the effects of the constrained LL and RR
insertions in $B_s \to K^{(*)-}K^{(*)+}$ decays are small because of
the absence of gluino mass enhancement. For
$m^2_{\tilde{g}}/m^2_{\tilde{q}}=9$,  the constrained LR insertion
can provide sizable contributions to all observables of $B_{s}\to
K^{(*)-}K^{(*)+}$ decays except $\mathcal{A}^{dir}_{CP}(B_{s}\to
K^{-}K^{+})$, and
 many observables are sensitive to the modulus and the phase of the LR insertion parameter.
   Near future experiments at Fermilab Tevatron and CERN LHC-b can test
these predictions and shrink/reveal the mass insertion parameter
spaces.
\end{abstract}

\vspace{0cm} \noindent {\bf PACS Numbers:  12.60.Jv,
  12.15.Ji, 12.38.Bx, 13.25.Hw}

\newpage
\section{Introduction}

The flavor changing neutral current (FCNC) processes in the $b\to s$
transition are sensitive to the effects of  New Physics (NP) beyond
the Standard Model (SM).
Recently, both CDF and D{\O} collaborations have announced their
measurements of extracted CP violating phase $\phi_s^{J/\psi\phi}$
associated with $B_s^0-\bar{B}_s^0$ mixing
\cite{Aaltonen:2007he,:2008fj,Tonelli:2008ey}. The CP violating
phase measured by both CDF and D{\O} is $\phi^{J/\psi\phi}_s \in
[0.20, 2.84]$ at 95\% C.L. \cite{exphis}, which is  much larger than
its SM value $\phi^{J/\psi\phi,\mbox{\scriptsize
SM}}_s=2\beta_s^{\mbox{\scriptsize
 SM}}\equiv2\mbox{arg}\left(-\frac{V_{ts}V^*_{tb}}{V_{cs}V^*_{cb}}\right)\approx
0.04$
\cite{Bona:2009tn,Lenz:2006hd,Silvestrini:2008zza,Bona:2008jn,HFAG2010}.
More recently, the D{\O} collaboration has reported evidence for an
anomalously large CP violation in the like-sign dimuon charge
asymmetry in semileptonic B-hadron decays
$A^b_{sl}=\left(-9.57\pm2.51 (\mbox{stat})\pm1.46
 (\mbox{syst})\right)\times10^{-3}$ \cite{Abazov:2010hj},
which differs by 3.2 standard deviations from the SM prediction
$A^{b,\mbox{\scriptsize SM}}_{sl} = (-2.3^{+0.5}_{-0.6})\times
10^{-4}$ \cite{Lenz:2006hd,Lenz:2008zza}.
Although the errors of the data are still large, these deviations
from the SM could be attributed to the presence of non-SM flavor
violationin the $b \to s,d$ nonleptonic decays.

Recently, the CDF collaboration  has made the first measurement of
charmless two-body $B_s \rightarrow K^-K^+$ decay, $\mathcal{B}(B_s
\rightarrow K^-K^+)=(26.5\pm4.4)\times10^{-6}$
\cite{Louvot:2009ie,Morello:2006pv,HFAG2010}. The measurement is
important for understanding  $B_{s}$ physics, and also implies that
many $B_{s}$ decay modes could be precisely measured at the LHC-b.
Comparing with  the  theoretical predictions for these observables
in Refs. \cite{Beneke:2003zv,Ali:2007ff,Williamson:2006hb}, one
would find that the experimental measurements of branching ratio are
in agreement with the SM predictions within their large theoretical
uncertainties. However, NP effects would be still possible to render
other observable deviated from the SM expectation with the branching
ratios nearly unaltered \cite{Baek:2006pb}.

Supersymmetry (SUSY) is an extension of the SM, which emerges as one
of the promising candidates for NP beyond the SM. In general SUSY, a
new source of flavor violation is introduced by the squark mass
matrices, which usually can not be diagonalized on the same basis as
the quark mass matrices. This means gluinos (and other gauginos)
will have flavor-changing couplings to quarks and squarks, which
implies the FCNCs could be mediated by gluinos and thus have strong
interaction strength. It is customary to rotate the effects so they
occur in squark propagators rather than in couplings, and to
parameterize them in terms of dimensionless mass insertion (MI)
parameters $(\delta^{u,d}_{AB})_{ij}$ with $(A,B)=(L,R)$ and
$(i,j=1,2,3)$.

$B^0_s-\bar{B}^0_s$ mixing, $B_{s}\to K^{(*)-}K^{(*)+}$ and $B\to
X_s\gamma$ decays are all induced by  the  $b \to s$  transition,
and they involve the same set of the MI parameters.  Inspired by the
 recent measurements from
 CDF and D{\O} collaborations , we study
$B^0_s-\bar{B}^0_s$ mixing, $B_{s}\to K^{(*)-}K^{(*)+}$ and $B\to
X_s\gamma$ decays in the usual MI approximation
\cite{Hall:1985dx,Gabbiani:1996hi} of general SUSY models, where
flavor violation due to the gluino mediation can be important. The
chargino-stop and the charged Higgs-top loop contributions are
parametrically suppressed relative to the gluino contributions, and
thus are ignored following
\cite{Gabbiani:1988rb,Hagelin:1992tc,Gabrielli:1995bd,Gabbiani:1996hi}.
 Following the similar way to our previous article \cite{Wang:2010zzh}, we consider the LL, RR, LR and RL four kinds of the MIs with
 $m^2_{\tilde{g}}/m^2_{\tilde{q}}=0.25,1,4,9$, respectively.
 We find that the LL and RR insertions for all cases of $m^2_{\tilde{g}}/m^2_{\tilde{q}}$ values
 as well as the LR insertion for $m^2_{\tilde{g}}/m^2_{\tilde{q}}=9$ case could explain current
 experimental data simultaneously.
 For $m^2_{\tilde{g}}/m^2_{\tilde{q}}=9$,  the constrained LR MI could significantly
affect all observables of $B_{s}\to K^{(*)-}K^{(*)+}$ decays except
$\mathcal{A}^{dir}_{CP}(B_{s}\to K^{-}K^{+})$
 without conflict with all related data at 95\% C.L.
While the constrained LL and RR insertions from $B^0_s-\bar{B}^0_s$
mixing have small effects in $B_{s}\to K^{(*)-}K^{(*)+}$ decays
because of the absence of gluino mass enhancement. Therefore, with
the ongoing $B$-physics at Tevatron, in particular with the onset of
the LHC-b experiment,
 we expect a wealth of  $B_s$ data and  measurements of
these observables  could restrict or reveal the parameter spaces of
 the LR (LL and RR) insertions in the near future.

The paper is arranged as follows.  In Sec. 2, the relevant formulas
for $B_{s}\to K^{(*)-}K^{(*)+}$ decays and $B^0_s-\bar{B}^0_s$
mixing are presented.  Sec. 3 deals with the numerical results.
Using our constrained  MI parameter spaces from $B_{s}\to
K^{(*)-}K^{(*)+}$ decay and $B^0_s-\bar{B}^0_s$ mixing, we explore
the MI effects on the
 other observable observables, which have not been measured yet in
$B_{s}\to K^{(*)-}K^{(*)+}$ decays.
 Sec. 4 contains our summary and conclusion. Theoretical input parameters are summarized
in the Appendix.

\section{The theoretical frame}

\subsection{ $B_s\to K^{(*)-}K^{(*)+}$ decays}
\label{BTOMM}
\subsubsection{The decay amplitudes in the SM}
  In the SM, the effective Hamiltonian for
  the $b\to s u\bar{u}$ transition at the scale $\mu\sim m_{b}$ is given by   \cite{Buchalla:1995vs}
 \begin{eqnarray}
 \mathcal{H}^{SM}_{eff}(\Delta B=1)=\frac{G_F}{\sqrt{2}}\sum_{p=u, c}
 \lambda_p \Biggl(C^{SM}_1Q_1^p+C^{SM}_2Q_2^p
 +\sum_{i=3}^{10}C^{SM}_iQ_i+C^{SM}_{7\gamma}Q_{7\gamma}
 +C^{SM}_{8g}Q_{8g} \Biggl)+ \mbox{h.c.},
 \label{HeffSM}
 \end{eqnarray}
where  $\lambda_p=V_{pb}V_{ps}^* $ with  $p\in \{u,c\}$ are
Cabibbo-Kobayashi-Maskawa (CKM) factors, the Wilson coefficients
within the SM  $C^{SM}_i$ can be found in Ref.
\cite{Buchalla:1995vs}, and the relevant operators $Q_i$ are given
as
    \begin{eqnarray}
    && \!\!\!\! \!\!\!\! \!\!\!\! \!\!\!\! \!\!\!\! \!\!\!\!
    Q^{p}_{1}=({\bar{p}}_{\alpha}\gamma^\mu Lb_{\alpha})
               ({\bar{s}}_{\beta}\gamma_\mu L p_{\beta} ),
    \ \ \ \ \ \ \ \ \ \ \ \ \ \
    Q^{p}_{2}=({\bar{p}}_{\alpha}\gamma^\mu Lb_{\beta} )
               ({\bar{s}}_{\beta} \gamma_\mu Lp_{\alpha}), \nonumber\\
     &&  \!\!\!\! \!\!\!\! \!\!\!\! \!\!\!\! \!\!\!\! \!\!\!\!
    Q_{3}=({\bar{s}}_{\alpha}\gamma^\mu Lb_{\alpha})\sum\limits_{q^{\prime}}
           ({\bar{q}}^{\prime}_{\beta} \gamma_\mu Lq^{\prime}_{\beta} ),
    \ \ \ \ \ \ \ \ \ \
    Q_{4}=({\bar{s}}_{\beta} \gamma^\mu Lb_{\alpha})\sum\limits_{q^{\prime}}
           ({\bar{q}}^{\prime}_{\alpha}\gamma_\mu Lq^{\prime}_{\beta} ),   \nonumber \\
    &&  \!\!\!\! \!\!\!\! \!\!\!\! \!\!\!\! \!\!\!\! \!\!\!\!
    Q_{5}=({\bar{s}}_{\alpha}\gamma^\mu Lb_{\alpha})\sum\limits_{q^{\prime}}
           ({\bar{q}}^{\prime}_{\beta} \gamma_\mu Rq^{\prime}_{\beta} ),
    \ \ \ \ \ \ \ \ \ \
    Q_{6}=({\bar{s}}_{\beta} \gamma^\mu Lb_{\alpha})\sum\limits_{q^{\prime}}
           ({\bar{q}}^{\prime}_{\alpha}\gamma_\mu Rq^{\prime}_{\beta} ), \nonumber\\
    &&  \!\!\!\! \!\!\!\! \!\!\!\! \!\!\!\! \!\!\!\! \!\!\!\!
    Q_{7}=\frac{3}{2}({\bar{s}}_{\alpha}\gamma^\mu Lb_{\alpha})
           \sum\limits_{q^{\prime}}e_{q^{\prime}}
           ({\bar{q}}^{\prime}_{\beta} \gamma_\mu Rq^{\prime}_{\beta} ),    \ \ \ \
    Q_{8}=\frac{3}{2}({\bar{s}}_{\beta} \gamma^\mu Lb_{\alpha})
           \sum\limits_{q^{\prime}}e_{q^{\prime}}
           ({\bar{q}}^{\prime}_{\alpha}\gamma_\mu Rq^{\prime}_{\beta} ), \nonumber\\
    &&  \!\!\!\! \!\!\!\! \!\!\!\! \!\!\!\! \!\!\!\! \!\!\!\!
    Q_{9}=\frac{3}{2}({\bar{s}}_{\alpha}\gamma^\mu Lb_{\alpha})
           \sum\limits_{q^{\prime}}e_{q^{\prime}}
           ({\bar{q}}^{\prime}_{\beta}\gamma_\mu L q^{\prime}_{\beta} ),
    \ \ \ \
    Q_{10}=\frac{3}{2}({\bar{s}}_{\beta} \gamma^\mu Lb_{\alpha})
           \sum\limits_{q^{\prime}}e_{q^{\prime}}
           ({\bar{q}}^{\prime}_{\alpha}\gamma_\mu Lq^{\prime}_{\beta} ), \nonumber\\
    &&\!\!\!\! \!\!\!\! \!\!\!\! \!\!\!\! \!\!\!\! \!\!\!\!
    Q_{7{\gamma}}=\frac{e}{8{\pi}^{2}}m_{b}{\bar{s}}_{\alpha}
           {\sigma}^{{\mu}{\nu}}R
            b_{\alpha}F_{{\mu}{\nu}},
     \ \ \ \ \ \ \ \ \ \ \ \ \
    Q_{8g}=\frac{g_s}{8{\pi}^{2}}m_{b}{\bar{s}}_{\alpha}
           {\sigma}^{{\mu}{\nu}}R
            T^{a}_{{\alpha}{\beta}}b_{\beta}G^{a}_{{\mu}{\nu}},
            \label{Eq:SMoperator}
    \end{eqnarray}
where $\alpha$ and $\beta$ are color indices, and $L(R)=(1\mp
\gamma_5)$.

 With the  effective Hamiltonian given in Eq.
(\ref{HeffSM}),  one can  write the decay amplitudes for the
relevant two-body hadronic
 $B\to M_{1}M_{2}$ decays as
\begin{eqnarray}
  \mathcal{A}^{SM}(B\to M_1M_2)&=&\left< M_1M_2|
  {\cal H}^{SM}_{eff}(\Delta B=1)|B \right> \nonumber\\
  &=&\sum_p \sum_i \lambda_p
  C^{SM}_i(\mu)\left<M_1M_2|Q_i(\mu)|B\right>.
  \end{eqnarray}
The essential theoretical difficulty for obtaining the decay
amplitude arises  from the  evaluation of hadronic matrix elements
$\langle M_1M_2|Q_i(\mu)|B\rangle$, for which we will employ the QCD
factorization (QCDF) \cite{BBNS} throughout  this paper.
 We will use the QCDF amplitudes of these decays derived in the comprehensive papers
 \cite{Beneke:2003zv,Beneke:2006hg} as  inputs for the SM expectations.

\subsubsection{SUSY effects in the decays}

\begin{figure}[t]
\begin{center}
\includegraphics[scale=1.15]{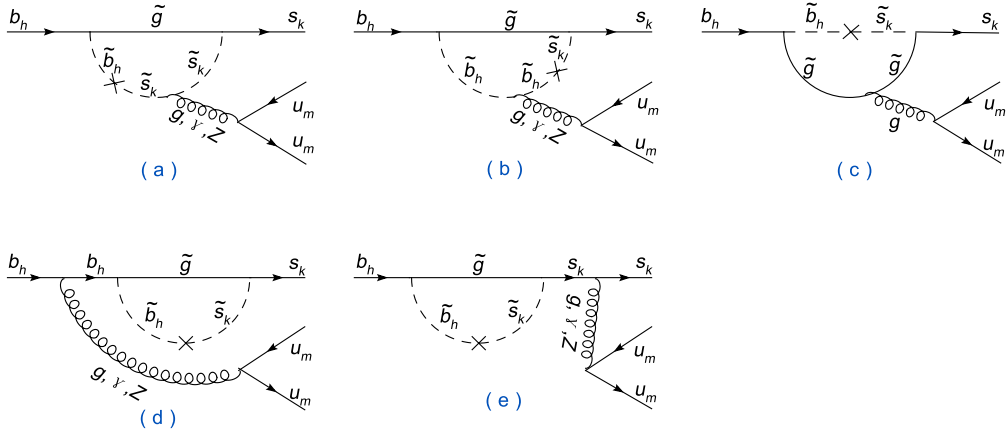}
\end{center}\vspace{-0.8cm}
\caption{ Penguin diagrams for $b\to su\bar{u}$ process with gluino
exchanges at the first order in mass insertion, where
$h,k,m={L,R}$.}
 \label{fig:penguinMIA}
\end{figure}

\begin{figure}[t]
\begin{center}
\includegraphics[scale=1.1]{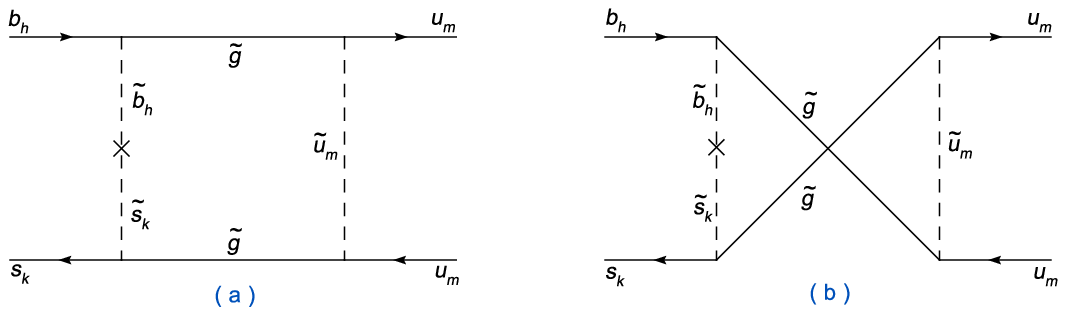}
\end{center}\vspace{-0.8cm}
\caption{ Box diagrams for $b\to su\bar{u}$ process with gluino
exchanges at the first order in mass insertion, where
$h,k,m={L,R}$.}
 \label{fig:boxMIA}
\end{figure}

In the SUSY extension of the SM with conserved R-parity, the
potentially most important contributions to Wilson coefficients of
penguin operators in the effective Hamiltonian arise from strong
penguin and box diagrams with gluino-squark loops. They contribute
to the FCNC processes because the gluinos have flavor-changing
coupling to the quark and squark eigenstates. In general SUSY, we
only consider these potentially large gluino box and penguin
contributions and neglect a multitude of other diagrams, which are
parametrically suppressed by small electroweak gauge coupling
\cite{Gabbiani:1988rb,Hagelin:1992tc,Gabrielli:1995bd,Gabbiani:1996hi}.
The relevant Wilson coefficients of the $b\to su\bar{u}$ process
due to the gluino penguin or box diagrams, which are shown in Fig.
\ref{fig:penguinMIA} and Fig. \ref{fig:boxMIA}, respectively,
involving the LL and LR insertions are given (at the scale $\mu \sim
m_W\sim m_{\tilde q}$) by
\cite{Gabbiani:1996hi,Baek:2001kc,Kane:2002sp,Ghosh:2002jpa}
\begin{eqnarray}
C_3^{SUSY}(m_{\tilde q})&=&-\frac{\alpha^2_s(m_{\tilde
q})}{2\sqrt{2}G_F\lambda_tm^2_{\tilde{q}}}
\left(-\frac{1}{9}B_1(x)-\frac{5}{9}B_2(x)-\frac{1}{18}P_1(x)-\frac{1}{2}P_2(x)\right)(\delta^d_{LL})_{23},\nonumber\\
C_4^{SUSY}(m_{\tilde q})&=&-\frac{\alpha^2_s(m_{\tilde
q})}{2\sqrt{2}G_F\lambda_tm^2_{\tilde{q}}}
\left(-\frac{7}{3}B_1(x)+\frac{1}{3}B_2(x)+\frac{1}{6}P_1(x)+\frac{3}{2}P_2(x)\right)(\delta^d_{LL})_{23},\nonumber\\
C_5^{SUSY}(m_{\tilde q})&=&-\frac{\alpha^2_s(m_{\tilde
q})}{2\sqrt{2}G_F\lambda_tm^2_{\tilde{q}}}
\left(\frac{10}{9}B_1(x)+\frac{1}{18}B_2(x)-\frac{1}{18}P_1(x)-\frac{1}{2}P_2(x)\right)(\delta^d_{LL})_{23},\nonumber\\
C_6^{SUSY}(m_{\tilde q})&=&-\frac{\alpha^2_s(m_{\tilde
q})}{2\sqrt{2}G_F\lambda_tm^2_{\tilde{q}}}
\left(-\frac{2}{3}B_1(x)+\frac{7}{6}B_2(x)+\frac{1}{6}P_1(x)+\frac{3}{2}P_2(x)\right)(\delta^d_{LL})_{23},\nonumber\\
C_{7\gamma}^{SUSY}(m_{\tilde q})&=&\frac{8\pi\alpha_s(m_{\tilde
q})}{9\sqrt{2}G_F\lambda_tm^2_{\tilde{q}}}
\left[(\delta^d_{LL})_{23}M_4(x)-(\delta^d_{LR})_{23}\left(\frac{m_{\tilde{g}}}{m_b}\right)4B_1(x)\right],\nonumber\\
C_{8g}^{SUSY}(m_{\tilde q})&=&-\frac{2\pi\alpha_s(m_{\tilde
q})}{\sqrt{2}G_F\lambda_tm^2_{\tilde{q}}}
\left[(\delta^d_{LL})_{23}\left(\frac{3}{2}M_3(x)-\frac{1}{6}M_4(x)\right)\right.\nonumber\\
&&\ \ \ \ \ \ \ \ \ \ \ \ \  \ \ \ \ \ \
\left.+(\delta^d_{LR})_{23}\left(\frac{m_{\tilde{g}}}{m_b}\right)\frac{1}{6}\left(4B_1(x)-9x^{-1}B_2(x)\right)\right],\label{Eq.SUSYWC}
\end{eqnarray}
where $x\equiv m^2_{\tilde{g}}/m^2_{\tilde{q}}$, and the loop
functions $B_i(x),P_i(x),M_i(x)$ can be found in Ref.
\cite{Baek:2001kc}. For the RR and RL insertions, we have additional
operators $\tilde{Q}_{i=3\ldots6,7\gamma,8g}$ that are obtained by
$L\leftrightarrow R$ in the SM operators  given in Eq.
(\ref{Eq:SMoperator}). The associated Wilson coefficients
$\widetilde{C}^{SUSY}_{i=3\ldots6,7\gamma,8g}$ are determined by the
expressions as above with the replacement $L\leftrightarrow R$. The
remaining coefficients are either dominated by their SM $(C_{1,2})$
or electroweak penguins $(C_{7\ldots10})$ and therefore small.

The  SUSY Wilson coefficients at low energy $C^{SUSY}_i(\mu\sim
m_b)$ can be obtained from $C^{SUSY}_i(m_{\tilde q})$ in Eq.
(\ref{Eq.SUSYWC})  by using the renormalization group equation as
discussed in Ref. \cite{Buchalla:1995vs}
\begin{eqnarray}
C(\mu)=U_5(\mu,m_{\tilde q})C(m_{\tilde q}),
\end{eqnarray}
where $C$ is the $6\times1$ column vector of the Wilson coefficients
and $U_5(\mu, m_{\tilde q})$  \cite{Buchalla:1995vs} is the
five-flavor $6\times6$ evolution matrix.
The coefficients $C^{SUSY}_{7\gamma}$ and $C^{SUSY}_{7g}$ at the
$\mu\sim m_b$ scale are given by  \cite{Buras:1999da,He:2001kn}
\begin{eqnarray}
C^{SUSY}_{7\gamma}(\mu)&=&\eta^2C^{SUSY}_{7\gamma}(m_{\tilde q})+\frac{8}{3}(\eta-\eta^2)C^{SUSY}_{8g}(m_{\tilde q}),\nonumber\\
C^{SUSY}_{8g}(\mu)&=&\eta C^{SUSY}_{8g}(m_{\tilde q}),
\end{eqnarray}
with $\eta=(\frac{\alpha_s(m_{\tilde
q})}{\alpha_s(m_t)})^{\frac{2}{21}}(\frac{\alpha_s(m_t)}{\alpha_s(m_b)})^{\frac{2}{23}}$.

\subsubsection{The total decay amplitudes}

For the LL and LR insertions, the NP effective operators have the
same chirality with the SM ones, so the total decays amplitudes can
be obtained from the SM ones in Refs.
\cite{Beneke:2003zv,Beneke:2006hg} by replacing
\begin{eqnarray}
C^{SM}_i\rightarrow C^{SM}_i+C^{SUSY}_i.
\end{eqnarray}
For the RL and RR insertions, the NP effective operators have the
opposite chirality with the SM ones, and we can get the
corresponding decay amplitudes from the SM decay amplitudes by
following replacements \cite{Kagan:2004ia}
\begin{eqnarray}
C^{SM}_i\rightarrow C^{SM}_i-\widetilde{C}^{SUSY}_i,
\end{eqnarray}
for $A(B_s\rightarrow K^-K^+)$  and $A_{0,\parallel}(B_s\rightarrow
K^{*-}K^{*+})$, as well as
\begin{eqnarray}
C^{SM}_i\rightarrow C^{SM}_i+\widetilde{C}^{SUSY}_i,
\end{eqnarray}
for $A(B_s\rightarrow K^{*-}K^+),$ $A(B_s\rightarrow K^-K^{*+})$,
and $A_{\perp}(B_s\rightarrow K^{*-}K^{*+})$.

Then the total branching ratio reads
\begin{eqnarray}
\mathcal{B}(B_{s}\to M_1 M_2)=\frac{\tau_{B_s} |p_c |}{8\pi
m_{B_s}^2}\left|\mathcal{A}(B_s\rightarrow M_1 M_2)\right|^2,
\end{eqnarray}
where $\tau_{B_{s}}$ is the $B_{s}$ lifetime, $|p_c|$ is the center
of mass momentum in the center of mass frame of $B_s$ meson.

In $B_s\to VV$ decay, the two vector mesons have the same helicity,
therefore three different polarization states are possible, one
longitudinal and two transverse, and we define the corresponding
helicity amplitudes as $\mathcal{A}_{0,\pm}$. Transverse
$(\mathcal{A}_{\parallel,\perp})$ and helicity $(\mathcal{A}_{\pm})$
amplitudes are related by
$\mathcal{A}_{\parallel,\perp}=\frac{\mathcal{A}_+\pm\mathcal{A}_-}{\sqrt{2}}$.
Then we have
\begin{eqnarray}
\left|\mathcal{A}(B_s\to
VV)\right|^2=|\mathcal{A}_0|^2+|\mathcal{A}_+|^2+|\mathcal{A}_-|^2
=|\mathcal{A}_0|^2+|\mathcal{A}_\parallel|^2+|\mathcal{A}_\perp|^2.
\end{eqnarray}
The longitudinal(transverse) polarization fractions $f_L$($f_\perp$)
are defined by
\begin{eqnarray}
f_{L,\perp}(B_s\to
VV)=\frac{\Gamma_{L,\perp}}{\Gamma}=\frac{|\mathcal{A}_{0,\perp}|^2}
{|\mathcal{A}_0|^2+|\mathcal{A}_\parallel|^2+|\mathcal{A}_\perp|^2}.
\end{eqnarray}

For the CP asymmetries (CPAs) of $B_s$ meson decays, there is an additional
complication due to $B^0_s-\bar{B}^0_s$ mixing. There are four cases
that one encounters for neutral $B_s$ decays, as discussed in Ref.
 \cite{Gronau:1989zb,Soto:1988hf,Palmer:1994ec,Ali:1998gb}:
\begin{itemize}
\item \textbf{Case (i)}:
$B^0_s\to f, \bar{B}^0_s\to \bar{f}$, where $f$ or $\bar{f}$ is not a
common final state of $B^0_s$ and $\bar{B}^0_s$, for example $B^0_s\to
K^-\pi^+,K^-\rho^+,K^{*-}\pi^+,K^{*-}\rho^+$.
\item \textbf{Case (ii)}:
$B^0_s\to (f=\bar{f})\leftarrow\bar{B}^0_s$ with $f^{CP}=\pm f$,
involving final states which are CP eigenstates, i.e., decays such
as $B^0_s\to K^-K^+$.
\item \textbf{Case (iii)}:
$B^0_s\to (f=\bar{f})\leftarrow\bar{B}^0_s$ with $f^{CP}\neq\pm f$,
involving final states which are not CP eigenstates. They include
decays such as $B^0_s\to VV$, as the $VV$ states are not  CP
eigenstates.
\item \textbf{Case (iv)}:
 $B^0_s\to (f\&\bar{f})\leftarrow \bar{B}^0_s$ with $f^{CP}\neq
f$, i.e., both $f$ and $\bar{f}$ are common final states of $B^0_s$
and $\bar{B}^0_s$, but they are not CP eigenstates. Decays
$B^0_s(\bar{B}^0_s)\to K^{*-}K^+,K^{-}K^{*+}$  belong to this case.
\end{itemize}

For case (i) decays, there is only  direct CPA
($\mathcal{A}_{CP}^{dir}$)  since no mixing is  involved for these
decays. For cases (ii) and (iii), their CPAs would involve
$B^0_s-\bar{B}^0_s$ mixing. The $\mathcal{A}_{CP}^{dir}$ and the
mixing-induced CPA ($\mathcal{A}_{CP}^{mix}$) are defined
as\footnote{ We use a similar sign convention to that of
\cite{Fleischer:2005vz} for self-tagging $B^0_s$ and charged $B$
decays.}
\begin{eqnarray}
\mathcal{A}_{CP}^{k,dir}(B^0_s\to
f)=\frac{\left|\lambda_k\right|^2-1}{\left|\lambda_k\right|^2+1},~~
\mathcal{A}_{CP}^{k,mix}(B^0_s\to
f)=\frac{2\mbox{Im}(\lambda_k)}{\left|\lambda_k\right|^2+1},
\end{eqnarray}
where  $k=0,\parallel,\perp$ for $B\to VV$ decays and $k=0$ for
$B\to PP,PV$ decays, in addition,
$\lambda_k=\frac{q}{p}\frac{\mathcal{A}_k(\overline{B}^0\rightarrow
\bar{f})}{\mathcal{A}_k(B^0_s\rightarrow f)}$ for CP case (i) and
$\lambda_k=\frac{q}{p}\frac{\mathcal{A}_k(\overline{B}^0\rightarrow
f)}{\mathcal{A}_k(B^0_s\rightarrow f)}$ for CP cases (ii) and (iii).

Case (iv) also involves mixing but requires additional formulas.
Here one studies the four time-dependent decay widths for $B^0_s(t)\to
f$, $\bar{B}^0_s(t)\to \bar{f}$, $B^0_s(t)\to \bar{f}$ and
$\bar{B}^0_s(t)\to f$
 \cite{Gronau:1989zb,Soto:1988hf,Palmer:1994ec,Ali:1998gb}. These
time-dependent widths can be expressed by four basic matrix elements
 \cite{Palmer:1994ec}
\begin{eqnarray} g&=&\langle
f|\mathcal{H}_{eff}|B^0_s\rangle,~~~~h=\langle
f|\mathcal{H}_{eff}|\bar{B}^0_s\rangle, \nonumber\\
\bar{g}&=&\langle
\bar{f}|\mathcal{H}_{eff}|\bar{B}^0_s\rangle,~~~\bar{h}=\langle
\bar{f}|\mathcal{H}_{eff}|B^0_s\rangle,
\end{eqnarray}
which determine the decay matrix elements of $B^0_s\to f\&\bar{f}$
and of $\bar{B}^0_s\to f\&\bar{f}$ at $t=0$. We will also study the
following observables
\begin{eqnarray}
&&\mathcal{A}_{CP}^{k,dir}(B^0_s\&\bar{B}^0_s\to
f)=\frac{\left|\lambda'_k\right|^2-1}{\left|\lambda'_k\right|^2+1},~~
\mathcal{A}_{CP}^{k,mix}(B^0_s\&\bar{B}^0_s\to
f)=\frac{2\mbox{Im}(\lambda'_k)}{\left|\lambda'_k\right|^2+1},\\
&&\mathcal{A}_{CP}^{k,dir}(B^0_s\&\bar{B}^0_s\to
\bar{f})=\frac{\left|\lambda''_k\right|^2-1}{\left|\lambda''_k\right|^2+1},~~
\mathcal{A}_{CP}^{k,mix}(B^0_s\&\bar{B}^0_s\to
\bar{f})=\frac{2\mbox{Im}(\lambda''_k)}{\left|\lambda''_k\right|^2+1},
\end{eqnarray}
with $\lambda'_k=\frac{q}{p}(h/g)$ and
$\lambda''_k=\frac{q}{p}(\bar{g}/\bar{h})$. The signature of CP
violation is $\Gamma(\bar{B}^0_s(t)\to \bar{f}) \neq \Gamma(B^0_s(t)\to
f)$ and $\Gamma(\bar{B}^0_s(t)\to f) \neq \Gamma(B^0_s(t)\to \bar{f})$,
which means that $\mathcal{A}_{CP}^{k,dir}(B^0_s\&\bar{B}^0_s\to f)$
$\neq$ $-\mathcal{A}_{CP}^{k,dir}(B^0_s\&\bar{B}^0_s\to \bar{f})$ and/or
$\mathcal{A}_{CP}^{k,mix}(B^0_s\&\bar{B}^0_s\to f)$ $\neq$
$-\mathcal{A}_{CP}^{k,mix}(B^0_s\&\bar{B}^0_s\to \bar{f})$.

\subsection{$B^0_s-\bar{B}^0_s$ mixing}
\label{SEC.Mixing}

 The most general $B^0_s-\bar{B}^0_s$ mixing is
described by the effective Hamiltonian \cite{Becirevic:2001jj}
\begin{eqnarray}
\mathcal{H}_{eff}(\Delta
B=2)=\sum^5_{i=1}C'_iQ'_i+\sum^3_{i=1}\widetilde{C}'_i\widetilde{Q}'_i+h.c.,\label{HB2eff}
\end{eqnarray}
with
\begin{eqnarray}
Q'_1&=&(\bar{s}\gamma^\mu P_Lb)_1(\bar{s}\gamma_\mu P_Lb)_1,\nonumber\\
Q'_2&=&(\bar{s} P_Lb)_1(\bar{s} P_Lb)_1,\nonumber\\
Q'_3&=&(\bar{s} P_Lb)_8(\bar{s} P_Lb)_8,\nonumber\\
Q'_4&=&(\bar{s} P_Lb)_1(\bar{s} P_Rb)_1,\nonumber\\
Q'_5&=&(\bar{s} P_Lb)_8(\bar{s} P_Rb)_8, \label{Q}
\end{eqnarray}
where $P_{L(R)}=(1\mp\gamma_5)/2$ and the operators
$\widetilde{Q}'_{1,2,3}$ are obtained from $Q'_{1,2,3}$ by the
exchange $L\leftrightarrow R$. The hadronic matrix elements, taking
into account for renormalization effects, are defined as
\begin{eqnarray}
\langle \bar{B}^0_s|
Q'_1(\mu)|B^0_s\rangle&=&\frac{2}{3}m^2_{B_s}f^2_{B_s}B_1(\mu),\nonumber\\
\langle \bar{B}^0_s|
Q'_2(\mu)|B^0_s\rangle&=&-\frac{5}{12}m^2_{B_s}f^2_{B_s}S_{B_s}B_2(\mu),\nonumber\\
\langle \bar{B}^0_s|
Q'_3(\mu)|B^0_s\rangle&=&\frac{1}{12}m^2_{B_s}f^2_{B_s}S_{B_s}B_3(\mu),\nonumber\\
\langle \bar{B}^0_s|
Q'_4(\mu)|B^0_s\rangle&=&\frac{1}{2}m^2_{B_s}f^2_{B_s}S_{B_s}B_4(\mu),\nonumber\\
\langle \bar{B}^0_s|
Q'_5(\mu)|B^0_s\rangle&=&\frac{1}{6}m^2_{B_s}f^2_{B_s}S_{B_s}B_5(\mu),\label{EM}
\end{eqnarray}
with
$S_{B_s}=\left(\frac{m_{B_s}}{\overline{m}_b(\overline{m}_b)+\overline{m}_s(\overline{m}_b)}\right)^2$.

The Wilson coefficients $C'_i$ receive contributions from both the
SM and the SUSY loops: $C'_i\equiv C'^{SM}_i+C'^{SUSY}_i$. In the
SM, the $t-W$ box diagram generates only contribution to the
operator $Q'_1$, and the corresponding Wilson coefficient
$C'^{SM}_1$ at the $m_b$ scale is \cite{Buchalla:1995vs}
\begin{eqnarray}
C'^{SM}_1(m_{b})=\frac{G_F^2}{4
\pi^2}m_W^2(V_{ts}V^*_{tb})^2\eta_{2B}S_0(x_t)[\alpha_s(m_b)]^{-6/23}
\left[1+\frac{\alpha_s(m_b)}{4\pi}J_5\right],
\end{eqnarray}
where $x_t=m^2_t/m^2_W$ and $\eta_{2B}$ is the QCD correction.

\begin{figure}[h]
\begin{center}
\includegraphics[scale=1.1]{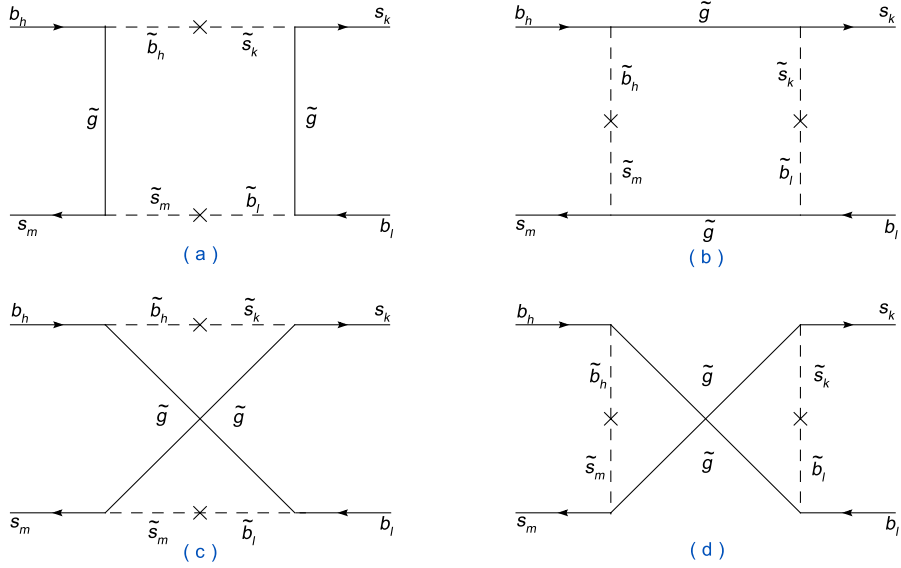}
\end{center}\vspace{-0.8cm}
\caption{ Feynman diagrams for $B^0_s-\bar{B}^0_s$ mixing in mass
insertion, where $h,k,l,m={L,R}$.}
 \label{fig:mixMIA}
\end{figure}
In general SUSY models, there are new contributions to
$B^0_s-\bar{B}^0_s$ mixing from the gluino-squark box diagrams,
which are shown in Fig. \ref{fig:mixMIA}, and the corresponding
Wilson coefficients $C'^{SUSY}_i$ (at the $m_{\tilde q}$ scale) are
given by
\cite{Gabbiani:1988rb,Hagelin:1992tc,Gabrielli:1995bd,Gabbiani:1996hi}
{\small \begin{eqnarray}
C'^{SUSY}_1(m_{\tilde q})&=&-\frac{\alpha_s^2}{216m^2_{\tilde{q}}}\left(24xf_6(x)+66\tilde{f}_6(x)\right)(\delta^d_{LL})^2_{23},\nonumber\\
C'^{SUSY}_2(m_{\tilde q})&=&-\frac{\alpha_s^2}{216m^2_{\tilde{q}}}204xf_6(x)(\delta^d_{RL})^2_{23},\nonumber\\
C'^{SUSY}_3(m_{\tilde q})&=&\frac{\alpha_s^2}{216m^2_{\tilde{q}}}36xf_6(x)(\delta^d_{RL})^2_{23},\nonumber\\
C'^{SUSY}_4(m_{\tilde
q})&=&-\frac{\alpha_s^2}{216m^2_{\tilde{q}}}\left[\left(504xf_6(x)-72\tilde{f}_6(x)\right)(\delta^d_{LL})_{23}(\delta^d_{RR})_{23}
-132\tilde{f}_6(x)(\delta^d_{LR})_{23}(\delta^d_{RL})_{23}\right],\nonumber\\
C'^{SUSY}_5(m_{\tilde
q})&=&-\frac{\alpha_s^2}{216m^2_{\tilde{q}}}\left[\left(24xf_6(x)+120\tilde{f}_6(x)\right)(\delta^d_{LL})_{23}(\delta^d_{RR})_{23}
-180\tilde{f}_6(x)(\delta^d_{LR})_{23}(\delta^d_{RL})_{23}\right].\label{mixCi}
\end{eqnarray}}
The loop functions $f_6(x)$ and $\tilde{f}_6(x)$ can be found in
Ref. \cite{Baek:2001kc}. Other Wilson coefficients
$\tilde{C}'^{SUSY}_{1,2,3}$ are obtained from $C'^{SUSY}_{1,2,3}$ by
exchange of $L\leftrightarrow R$.

The SUSY Wilson coefficients at the $m_b$ scale $C^{SUSY}_i( m_b)$
can be obtained by
\begin{eqnarray}
C_r(m_b)=\sum_i\sum_s\left(b^{(r,s)}_i+\eta'
c^{(r,s)}_i\right)\eta'^{a_i}C_s(m_{\tilde{q}}),
\end{eqnarray}
where $\eta'=\alpha_s(m_{\tilde{q}})/\alpha_s(m_t)$.  The magic
number $a_i$,  $b^{(r,s)}_i$ and $c^{(r,s)}_i$ can be found in Ref.
\cite{Becirevic:2001jj}. Renormalization group evolution of
$\tilde{C}_{1,2,3}$ can be  done in the same way as for $C_{1,2,3}$.

In terms of the effective Hamiltonian in Eq. (\ref{HB2eff}), the
mixing amplitude $M_{12}$ reads
\begin{eqnarray}
M_{12}=\frac{\langle B^0_s|\mathcal{H}_{eff}(\Delta
B=2)|\bar{B}^0_s\rangle}{2m_{B_s}}.
\end{eqnarray}

In the SM, the off-diagonal element of the decay
  matrix $\Gamma^{s,SM}_{12}$  may be written as \cite{Benekembs}
\begin{eqnarray}
\Gamma^{s,SM}_{12}=-\frac{G^2_Fm^2_b}{8\pi
M_{B_s}}(V_{cs}V^*_{cb})^2\left[G(x_c)\langle
B^0_s|Q_1|\bar{B}^0_s\rangle+G_2(x_c)\langle
B^0_s|Q_2|\bar{B}^0_s\rangle+\sqrt{1-4x_c}\hat{\delta}_{1/m}\right],
\end{eqnarray}
where $x_c=m_c^2/m_b^2$, $G(x_c)=0.030$, and $G_2(x_c)=-0.937$ at
the $m_b$ scale \cite{Benekembs}.  The $1/m_b$ corrections
$\hat{\delta}_{1/m}$ are given in Ref. \cite{1/mbcorrections}, and
$1/m^2_b$ corrections are not considered since they are small
\cite{Badin:2007bv}. It is important to note that, SUSY
contributions can significantly affect $M^s_{12}$, but have little
effect on $\Gamma^s_{12}$ which is dominated by the CKM-favored
$b\to sc\bar{c}$ tree-level decays, hence
$\Gamma^s_{12}=\Gamma^{s,SM}_{12}$ holds as a good approximation
\cite{Gamma12,He:2010fz,Lenz:2006hd}.

In general, the relevant CP violating phase between the
$B^0_s-\bar{B}^0_s$ amplitude and the amplitudes of the subsequent
$B^0_s$ and $\bar{B}^0_s$ decay to a common final state  could be
expressed as \cite{Lenz:2007nk}
\begin{eqnarray}
\phi_s=\mbox{arg}\left(-\frac{M^s_{12}}{\Gamma^s_{12}}\right).
\end{eqnarray}
The SM prediction for this phase is tiny,
$\phi^{\mbox{\scriptsize  SM}}_s\approx0.004$ \cite{Lenz:2006hd}.
The same additional contribution $\phi^{\mbox{\scriptsize  NP}}_s$
due to NP would change this observed phase,  i.e.,
$\phi_s=\phi^{\mbox{\scriptsize SM}}_s+\phi^{\mbox{\scriptsize
NP}}_s$.
In case of sizable
NP contributions,
 the following
approximation is used:
$\phi^{J/\psi\phi}_s\approx\phi_s\approx\phi^{\mbox{\scriptsize
NP}}_s$.

In this work, besides the CP violating phase $\phi_s^{J/\psi\phi}$,
the experimental bounds of the following observables will be
considered:
\begin{itemize}

\item  the $B_s$ mass difference:
$ \Delta M_s= 2\left|M_{12}^{s}\right|;$

\item the $B_s$ width difference \cite{Grossman}:
$\Delta\Gamma_s=\frac{4|Re(M^s_{12}\Gamma^{s*}_{12})|}{\Delta M_s}
\approx2|\Gamma^s_{12}|\mbox{cos}\phi_s;$

\item
the semileptonic CP asymmetry in $B_s$ decays \cite{ASL1,ASL2}:
$A^s_{SL}=\mbox{Im}\left(\frac{\Gamma^s_{12}}{M^s_{12}}\right)
=\frac{\Delta \Gamma_s}{\Delta M_s}~\mbox{tan}\phi_s.$

\end{itemize}

\section{Numerical results and analysis}
Now we are ready to present our numerical results and analysis.
First, we will show our estimations in the SM with the theoretical
input parameters listed in Table \ref{Tab.input} of Appendix. Then,
we will consider the SUSY effects with LL, RR, LR, and RL four kinds
of the MIs and constrain the relevant MI parameters with the
experimental data of $B_s\to K^{-}K^{+}$, $B\to X_s\gamma$ and
$B^0_s-\bar{B}^0_s$ mixing. In each of the MI scenarios to be
discussed, we will vary the MIs over the range
$|(\delta^d_{AB})_{23}|\leq 1$ to fully map the parameter space. We
will consider the weak phases resided in the complex MI parameters
$(\delta^d_{AB})_{23}$ and appeared in the SUSY Wilson coefficients
in Eq. (\ref{Eq.SUSYWC}) and Eq. (\ref{mixCi}), and these weak
phases are odd under a CP transformation.  Using the constrained
parameter spaces, we will give the MI SUSY predictions for the
branching ratios, the CPAs
 and the polarization fractions,
which have not been measured yet in $B_s\to K^{(*)-}K^{(*)+}$
decays.


The numerical results in the SM are presented in second column of
Table \ref{Tab.SUSYpredictions}. For the decays, the detailed error
estimations corresponding to the different types of theoretical
uncertainties have been already studied  in Refs.
 \cite{Beneke:2003zv,Beneke:2006hg,Xu:2009we}, and our SM results
 are consistent
with the ones in Refs. \cite{Beneke:2003zv,Beneke:2006hg,Xu:2009we}.
For $B^0_s-\bar{B}^0_s$ mixing, $\phi_s^{J/\psi\phi}$ and $A^s_{SL}$
are precisely predicted in the SM, and the uncertainties of $\Delta
M_s$ and $\Delta\Gamma_s$ mainly arise from the nonperturbative
quantity $f_{B_s}\sqrt{\hat{B}_{B_s}}$ and the CKM matrix elements.

\begin{table}[hbt]
\caption{The theoretical predictions for  $B_s\to K^{(*)-} K^{(*)+}$
decays and $B^0_s-\bar{B}^0_s$ mixing based on general SUSY models
with LR MI and $x=9$. $\mathcal{B}$  and $A^s_{SL}$
are in units of $10^{-6}$ and $10^{-2}$, respectively. The
corresponding SM predictions and relevant experimental data  are also listed for comparison.  }
\begin{center}{\footnotesize
\begin{tabular}
{l|ccc}\hline\hline
 Observables &Experimental ranges &SM predictions & SUSY values with \\
 &at 95\% C.L. &&$(\delta_{LR}^d)_{23}$ for $x=9$\\\hline
$\Delta M_s$&$[17.53,18.01]$&$[13.66,24.82]$&$[17.53,18.01]$\\\hline
$\Delta \Gamma_s$&$[0.05,0.33]$&$[0.10,0.21]$&$[0.10,0.21]$\\\hline
$\phi_s$&$[0.16,2.84]$&$[0.034,0.038]$&$[0.16,0.52]$\\\hline
$A^s_{SL}$&$[-0.04,2.96]$&$[0.02,0.05]$&$[0.11,0.46]$\\\hline

$\mathcal{B}(B_s\rightarrow
K^{-}K^+)$&$[17.70,35.30]$&$[9.20,45.52]$&$[22.80,35.30]$\\\hline
$\mathcal{B}(B_s\rightarrow
K^{*-}K^+)$&&$[2.56,23.19]$&$[2.39,5.78]$\\\hline
$\mathcal{B}(B_s\rightarrow
K^{-}K^{*+})$&&$[1.92,6.72]$&$[9.73,20.64]$\\\hline
$\mathcal{B}(B_s\rightarrow
K^{*-}K^{*+})$&&$[3.56,18.76]$&$[11.00,45.40]$\\\hline
$\mathcal{A}^{mix}_{CP}(B_s\rightarrow
K^{-}K^+)$&&$[0.25,0.49]$&$[0.52,0.79]$\\\hline
$\mathcal{A}^{mix}_{CP}(B_s\&\bar{B}_s\rightarrow
K^{*-}K^{+})$&&$[-0.34,0.07]$&$[-0.09,0.64]$\\\hline
$\mathcal{A}^{mix}_{CP}(B_s\&\bar{B}_s\rightarrow
K^{-}K^{*+})$&&$[-0.44,0.05]$&$[-0.14,0.61]$\\\hline
$\mathcal{A}^{L,mix}_{CP}(B_s\rightarrow
K^{*-}K^{*+})$&&$[0.70,0.95]$&$[0.79,0.98]$\\\hline
$\mathcal{A}^{dir}_{CP}(B_s\rightarrow
K^{-}K^+)$&&$[0.00,0.06]$&$[0.00,0.07]$\\\hline
$\mathcal{A}^{dir}_{CP}(B_s\rightarrow
K^{*-}K^{+})$&&$[-0.08,0.02]$&$[-0.17,0.00]$\\\hline
$\mathcal{A}^{dir}_{CP}(B_s\rightarrow
K^{-}K^{*+})$&&$[-0.10,0.10]$&$[-0.04,0.05]$\\\hline
$\mathcal{A}^{dir}_{CP}(B_s\&\bar{B}_s\rightarrow
K^{*-}K^{+})$&&$[-0.77,0.27]$&$[0.05,0.76]$\\\hline
$\mathcal{A}^{dir}_{CP}(B_s\&\bar{B}_s\rightarrow
K^{-}K^{*+})$&&$[-0.24,0.76]$&$[-0.76,-0.09]$\\\hline
$\mathcal{A}^{L,dir}_{CP}(B_s\rightarrow
K^{*-}K^{*+})$&&$[-0.13,0.21]$&$[-0.04,0.10]$\\\hline
$f_{L}(B_s\rightarrow
K^{*-}K^{*+})$&&$[0.36,0.88]$&$[0.75,0.96]$\\\hline
$f_{\bot}(B_s\rightarrow
K^{*-}K^{*+})$&&$[0.06,0.33]$&$[0.02,0.13]$\\\hline\hline
\end{tabular}}
\end{center}\label{Tab.SUSYpredictions}
\end{table}

Now we turn to the gluino-mediated SUSY contributions to  $B_s\to
K^{(*)-}K^{(*)+}$  decays and $B^0_s-\bar{B}^0_s$ mixing in the
framework of the MI approximation. The following experimental data
 will be used to constrain relevant MI
couplings \cite{exphis,Aaltonen:2007he,:2008fj,Abazov:2010hj}
\begin{eqnarray}
\phi^{J/\psi\phi}_s &\in& [0.20, 2.84]~ (\mbox{at}~ 95\%~\mbox{C.L.}),\label{exp.phis}\\
\Delta M_s&=&17.77\pm0.12,\label{exp.Dms}\\
 \Delta\Gamma_s&=&0.19\pm0.07,\label{exp.Dgamma}\\
A^s_{SL}&=&(1.46\pm0.75)\times10^{-2},\label{exp.As}\\
\mathcal{B}(B_s \rightarrow
K^-K^+)&=&(26.5\pm4.4)\times10^{-6}.\label{exp.Br}
\end{eqnarray}
In addition, the same set of the MI parameters also contribute to
$B\to X_s\gamma$, which the gluino-mediated contribution  can be
found in Ref. \cite{Kane:2002sp}.
Since the experimental measurement of $\mathcal{B}(B\to X_s\gamma)$
is in good agreement with the SM expectation,
this implies very stringent constraints on NP models. We will also
use \cite{HFAG2010}
\begin{eqnarray}
\mathcal{B}(B\to
X_s\gamma)=(3.55\pm0.24\pm0.09)\times10^{-4}\label{exp.Brsr}
\end{eqnarray}
to constrain the relevant MI parameters. Noted that above
experimental data at 95\% C.L. will be used to constrain the MI
parameters.

\subsection{LL insertion}
Let us first consider the LL insertion. The effects of the
 LL insertions in $B_s\to K^{(*)-}K^{(*)+}$ decays are almost
negligible because there is no the gluino mass enhancement, and
$\mathcal{B}(B_s \rightarrow K^-K^+)$ given in Eq. (\ref{exp.Br})
can not provide any useful constraint on $(\delta_{LL}^d)_{23}$. The
bound from $A^s_{SL}$ is weaker than one from $\phi^{J/\psi}_s$,
therefore $A^s_{SL}$ also does not give any useful constraint when
we consider all experimental data given in
Eqs.(\ref{exp.phis}-\ref{exp.Brsr}) to constrain four kinds of the
MI parameters.  So we only impose the experimental constraints of
$B^0_s-\bar{B}^0_s$ mixing and $B\to X_s\gamma$ decay, which are
shown in Eqs. (\ref{exp.phis}-\ref{exp.Dgamma}) and Eq.
(\ref{exp.Brsr}), respectively,  to restrict $(\delta^d_{LL})_{23}$.
\begin{figure}[h]
\begin{center}
\includegraphics[scale=0.72]{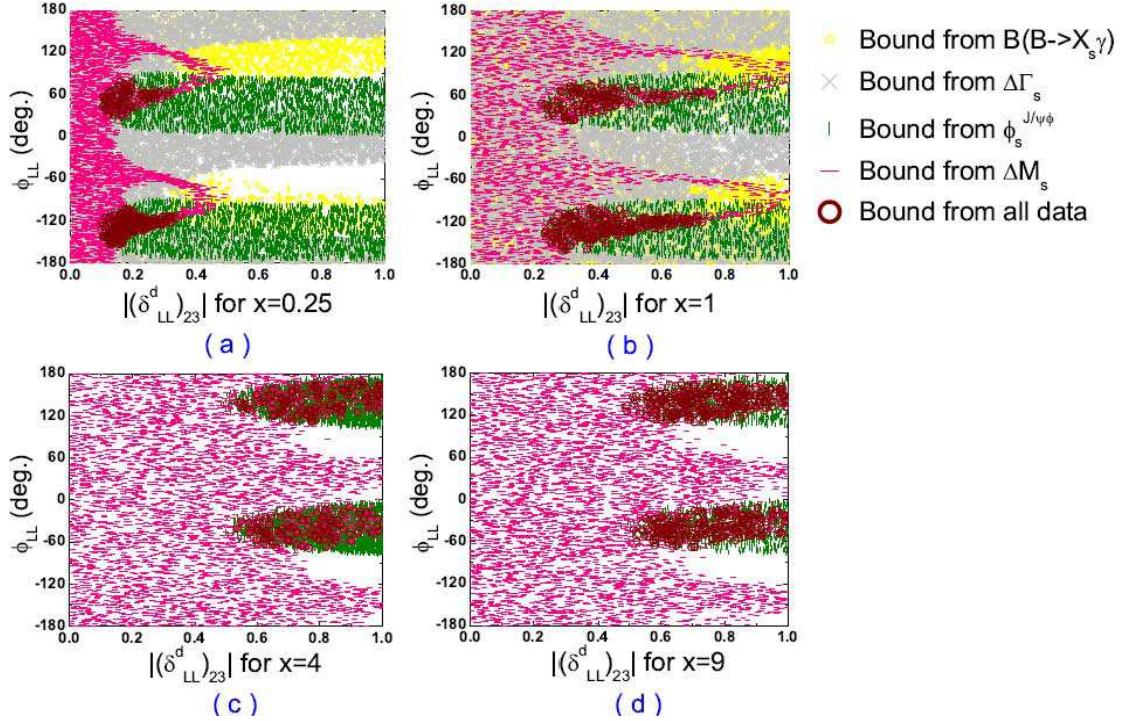}
\end{center}\vspace{-0.8cm}
\caption{ The allowed parameter spaces of the LL MI parameter
constrained by $B_s\to K^{-}K^{+}$, $B\to X_s\gamma$ and
$B^0_s-\bar{B}^0_s$ mixing at 95\% C.L. for the squark mass
$m_{\tilde{q}}=500$ GeV and the different values of $x$, and
$\phi_{LL}$ denotes the mixing parameters weak phase. }
 \label{fig:boundsLL}
\end{figure}

The constrained spaces of $(\delta_{LL}^d)_{23}$ for
$m_{\tilde{q}}=500$ GeV and different $x$ values  are demonstrated
in Fig. \ref{fig:boundsLL}, where the allowed parameter space for
the MI is shown as dictated by the constraints imposed by
$\mathcal{B}(B\to X_s\gamma)$ (yellow), $\Delta\Gamma_s$ (light
gray), $\phi^{J/\psi\phi}_s$ (olive) and $\Delta M_s$ (pink). The
wine region shows the allowed regions under the combined constraints
of $\mathcal{B}(B\to X_s\gamma)$, $\mathcal{B}(B_s \rightarrow
K^-K^+)$, $\Delta\Gamma_s$, $\Delta M_s$, $A^s_{SL}$ and
$\phi^{J/\psi\phi}_s$. From Fig. \ref{fig:boundsLL}, we see that the
constrained regions are very sensitive to the values of $x$. For
$x=0.25,1$, as shown in Fig. \ref{fig:boundsLL}(a-b),  the common
allowed regions  are constrained by $\Delta\Gamma_s$,
$\phi^{J/\psi\phi}_s$ and $\Delta M_s$, nevertheless
$\mathcal{B}(B\to X_s\gamma)$ does not give any further constraint.
For $x=4,9$, we don't show the constraints from $\mathcal{B}(B\to
X_s\gamma)$ in Fig. \ref{fig:boundsLL}(c-d) since the whole region
of $|(\delta^d_{LL})_{23}|\leq 1$ is allowed by the constraint of
$\mathcal{B}(B\to X_s\gamma)$. As displayed in Fig.
\ref{fig:boundsLL}(c-d),  the common allowed regions for $x=4$ and
$9$ cases  are constrained by $\phi^{J/\psi\phi}_s$ and $\Delta
M_s$, while  $\Delta\Gamma_s$ does not give any further constraint.
It is worth noting that,  for $x=0.25,1,4,9$, the lower limit of
$|(\delta^d_{LL})_{23}|$ is also constrained by
$\phi^{J/\psi\phi}_s$ since its data  are not consistent with its SM
value at 95\% C.L. The relevant numerical bounds on
$|(\delta^d_{LL})_{23}|$ with different $x$ values are summarized in
Table \ref{Tab.boundsLL}.
\begin{table}[t]
\caption{Bounds on the LL MI parameters from the measurements of $\mathcal{B}(B\to X_s\gamma)$, $\mathcal{B}(B_s
\rightarrow K^-K^+)$, $\Delta\Gamma_s$, $\Delta M_s$, $A^s_{SL}$ and
$\phi^{J/\psi\phi}_s$
at 95\% C.L. for the squark mass $m_{\tilde{q}}=500$ GeV. }
\begin{center}
\begin{tabular}
{c|cccc}\hline\hline $x$& $0.25$& $1$&$4$& $9$\\\hline
$|(\delta_{LL}^d)_{23}|$&$[0.10,0.35]$&$[0.22,0.76]$&$[0.54,1.00]$&$[0.49,1.00]$\\
$\phi_{LL}$(deg.)&$^{[29,79]}_{[-154,-102]}$&$^{[24,76]}_{[-162,-101]}$&$^{[112,170]}_{[-69,-10]}$&$^{[109,167]}_{[-68,-12]}$\\\hline
\end{tabular}
\end{center}\label{Tab.boundsLL}
\end{table}

In Ref. \cite{Altmannshofer:2009ne}, the constraint
$|(\delta_{LL}^d)_{23}|\leq0.5$ for
$m_{\tilde{g}},m_{\tilde{q}}\leq600$ GeV are derived from
$\mathcal{B}(B\to X_s\gamma)$ and $\mathcal{B}(B\to
X_s\ell^+\ell^-)$.
Compared with the existed bound in \cite{Altmannshofer:2009ne}, for
$x=0.25,1$, our upper limits of $|(\delta_{LL}^d)_{23}|$ are at the
same order as the previous ones, while the lower limits of
$|(\delta_{LL}^d)_{23}|$ are also given by $\phi^{J/\psi\phi}_s$ at
95\% C.L.. However, for $x=4,9$, our bounds on
$|(\delta_{LL}^d)_{23}|$ are greater than ones in Ref.
\cite{Altmannshofer:2009ne}.  Moreover, the bounds on the LL
insertion with small tan$\beta$  by $\Delta M_s$,
$\phi^{J/\psi\phi}_s$, $\mathcal{B}(B\to X_s\gamma)$, $A^{b\to
s\gamma}_{CP}$ and $S^{\phi K}_{CP}$  are also analyzed in detail in
Ref. \cite{Ko:2008xb}, for $m_{\tilde{q}}=500$ GeV and $x=1$ case,
$|(\delta_{LL}^d)_{23}|$ lies in $[0.42,0.44]\cup[0.90,0.95]$ with
tan$\beta=3$ and lies in $[0.40,0.65]$ with tan$\beta=10$.

The constrained LL MI  shown in Fig. \ref{fig:boundsLL} allows that
the theoretical prediction of $\Delta M_s$ lies in its 95\% C.L.
experimental range $[17.53,18.01]$ for $x=0.25,1,4,9$. However, the
ranges of $\phi^{J/\psi\phi}_s,~\Delta \Gamma_s$ and $A^s_{SL}$ are
narrower than their 95\% C.L. experimental ranges. For $x=0.25,1$,
the constrained LL insertion coupling allows
$\phi^{J/\psi\phi}_s\in[0.16,1.26]$, $\Delta \Gamma_s\in[0.05,0.20]$
and $A^s_{SL}\in[0.10,1.00]$.   For $x=4,9$, this coupling allows
$\phi^{J/\psi\phi}_s\in[0.16,0.52]$, $\Delta \Gamma_s\in[0.10,0.20]$
and $A^s_{SL}\in[0.10,0.47]$ .

Furthermore, we also explore the LL insertion effects in $B_{s}\to
K^{(*)-}K^{(*)+}$ decays.  After satisfying all experimental data at
95\% C.L. given in Eqs. (\ref{exp.phis}-\ref{exp.Brsr}),  the
constrained LL insertion  will not provide significant contribution
to $B_{s}\to K^{(*)-}K^{(*)+}$ decays. We find the upper limits of
$\mathcal{B}(B_{s}\to K^{*-}K^{*+})$,
$\mathcal{A}^{mix}_{CP}(B_s\&\bar{B}_s\rightarrow
K^{*-}K^{+},K^{-}K^{*+})$, $\mathcal{A}^{dir}_{CP}(B_s\rightarrow
K^{*-}K^{+})$ and $f_\perp(B_{s}\to K^{*-}K^{*+})$ are slightly
decreased from their SM ranges by the constrained LL insertion. The
lower limits of $\mathcal{A}^{L,dir}_{CP}(B_s\rightarrow
K^{*-}K^{*+})$ and $f_L(B_{s}\to K^{*-}K^{*+})$ are slightly
increased from their SM ranges by the constrained LL insertion. The
allowed range of $\mathcal{A}^{L,mix}_{CP}(B_s\rightarrow
K^{*-}K^{*+})$ is increased from its SM prediction $[0.70,0.95]$ to
[0.74,1.00] for $x=0.25$, [0.77,0.99] for $x=1$,  [0.81,0.97] for
$x=4$ and  [0.76,0.97] for $x=9$, respectively,  by the constrained
LL insertion. While all observables of $B_{s}\to K^{(*)-}K^{(*)+}$
decays are insensitive to the modulus and weak phase of
$(\delta_{LL}^d)_{23}$.

\subsection{RR insertion}
\begin{figure}[t]
\begin{center}
\includegraphics[scale=0.72]{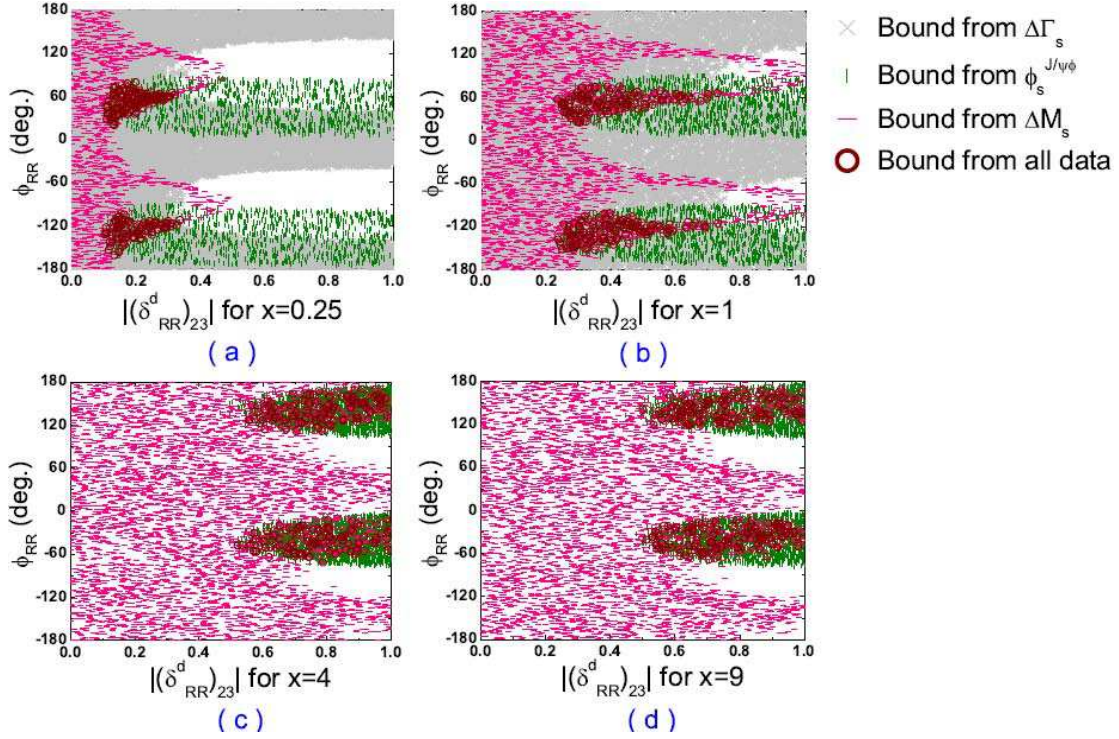}
\end{center}\vspace{-0.8cm}
\caption{ The allowed parameter spaces of the RR  MI parameters
constrained by $B_s\to K^{-}K^{+}$ decay and $B^0_s-\bar{B}^0_s$
mixing at 95\% C.L. for the squark mass $m_{\tilde{q}}=500$ GeV and
 the different values of $x$.}
 \label{fig:boundsRR}
\end{figure}

 For $B\to X_s\gamma$ decay, the situation of the RR
insertion is very different from the LL one since the related NP
amplitude (arising from right-handed currents) does not interfere
with the SM one. Moreover, the effects of the RR insertion in
$B_s\to K^-K^+$ are almost negligible also because of lacking the
gluino mass enhancement in the decay. Therefore
$(\delta_{RR}^d)_{23}$
 is strongly constrained by $B^0_s-\bar{B}^0_s$ mixing.
The constrained spaces of $(\sigma_{RR}^d)_{23}$ by
$B^0_s-\bar{B}^0_s$ mixing for $m_{\tilde{q}}=500$ GeV and different
$x$ values are demonstrated in Fig. \ref{fig:boundsRR}, and the
corresponding numerical ranges are summarized in Table
\ref{Tab.boundsRR}. From Fig. \ref{fig:boundsRR} and Table
\ref{Tab.boundsRR}, we can see that the allowed moduli and the
allowed phase ranges of the RR parameters are also very sensitive to
the values of $x$.

\begin{table}[t]
\caption{Bounds on the RR MI parameters from the measurements of
$B_s\to K^{-}K^{+}$, $B\to X_s\gamma$ and $B^0_s-\bar{B}^0_s$ mixing
at 95\% C.L. for the squark mass $m_{\tilde{q}}=500$ GeV. }
\begin{center}
\begin{tabular}
{c|cccc}\hline\hline $x$& $0.25$& $1$&$4$& $9$\\\hline
$|(\delta_{RR}^d)_{23}|$&$[0.10,0.34]$&$[0.23,0.73]$&$[0.52,1.00]$&$[0.50,1.00]$\\
$\phi_{RR}$(deg.)&$^{[20,86]}_{[-160,-104]}$&$^{[25,75]}_{[-153,-101]}$&$^{[111,170]}_{[-71,-11]}$&$^{[118,170]}_{[-69,-13]}$\\\hline
\end{tabular}
\end{center}\label{Tab.boundsRR}
\end{table}

The bound of $(\delta_{RR}^d)_{23}$   has been  obtained  in Refs.
\cite{Altmannshofer:2009ne,Ko:2008xb}. The contributions of the
product $(\delta_{LL}^d)_{23}(\delta_{RR}^d)_{23}$ are also
considered in Ref. \cite{Altmannshofer:2009ne}, and they obtain
$|(\delta_{RR}^d)_{23}|\leq0.8$ for
$m_{\tilde{g}},m_{\tilde{q}}\leq600$ GeV from $\mathcal{B}(B\to
X_s\gamma)$ and $\mathcal{B}(B\to X_s\ell^+\ell^-)$.
In Ref. \cite{Ko:2008xb}, the bounds on the RR insertion with small
tan$\beta$  from $\Delta M_s$, $\phi^{J/\psi\phi}_s$,
$\mathcal{B}(B\to X_s\gamma)$, $A^{b\to s\gamma}_{CP}$ and $S^{\phi
K}_{CP}$  are also analyzed in detail, for $m_{\tilde{q}}=500$ GeV
and $x=1$ case, $|(\delta_{RR}^d)_{23}|$ lies in $[0.36,0.69]$ when
tan$\beta=3$, and there is no common range when tan$\beta=10$.

The constrained RR insertion has the similar effects as the LL
insertion on the observables of $B_{s}\to K^{(*)-}K^{(*)+}$ decays
and $B^0_s-\bar{B}^0_s$ mixing, and we will not show them here.

\subsection{LR insertion}
The effect of the LR insertion  is very different from that of
either LL or RR. In these decays, the LR MI only generates
(chromo)magnetic operators $Q_{7\gamma,8g}$ and
$\tilde{Q}_{7\gamma,8g}$. Especially, the LR insertion is more
strongly constrained, since their contributions are enhanced by
$m_{\tilde{g}}/m_b$ due to the chirality flip from the gluino in the
loop. Thus, even a small $(\delta_{LR}^d)_{13}$ can have large
effects in $B\to X_s\gamma$ and $B_{s}\to K^{(*)-}K^{(*)+}$ decays.

\begin{figure}[b]
\begin{center}
\includegraphics[scale=0.72]{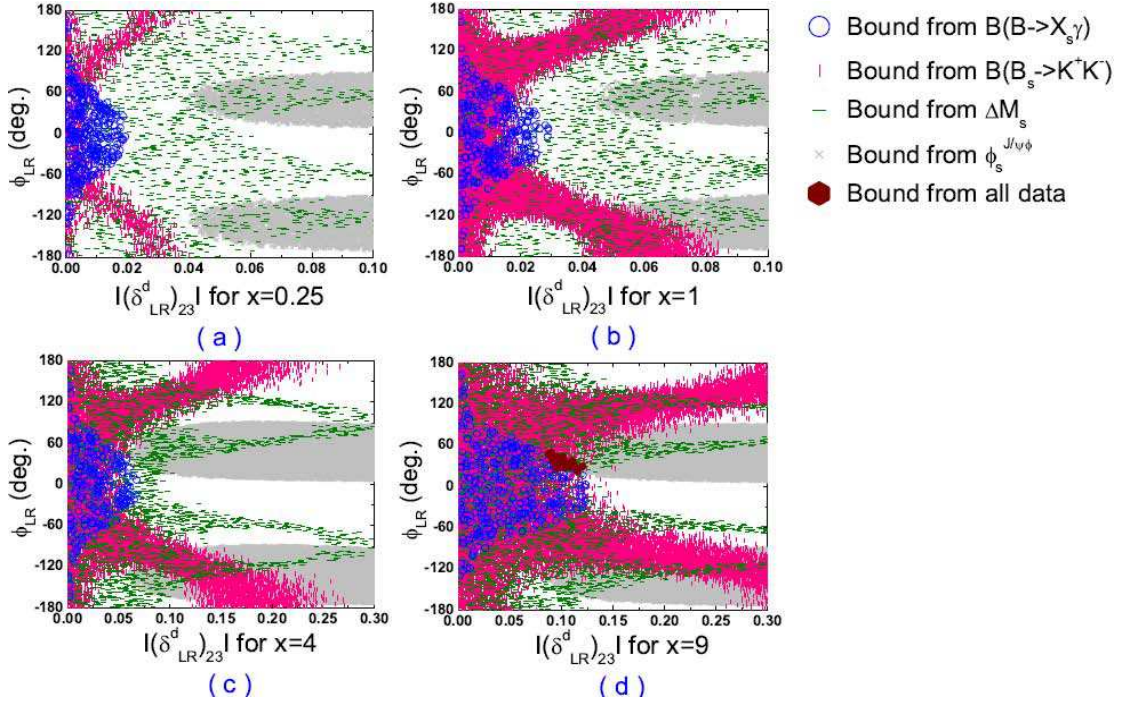}
\end{center}\vspace{-0.8cm}
\caption{ The allowed parameter spaces of the LR  MI parameters
constrained by $B_s\to K^{-}K^{+}$ decay and $B^0_s-\bar{B}^0_s$
mixing at 95\% C.L. for the squark mass $m_{\tilde{q}}=500$ GeV and
for the different values of $x$.}
 \label{fig:boundsLR}
\end{figure}
The constrained spaces of $(\delta_{LR}^d)_{23}$ from
$\mathcal{B}(B\to X_s\gamma)$, $\mathcal{B}(B_{s}\to K^{-}K^{+})$
and $B^0_s-\bar{B}^0_s$ mixing for $m_{\tilde{q}}=500$ GeV as well
as different $x$ are demonstrated in Fig. \ref{fig:boundsLR}.
$\Delta\Gamma_s$ cannot provide any further constraint on
$(\delta_{LR}^d)_{23}$ and we will not  show them in Fig.
\ref{fig:boundsLR}.
From the figure, we can see that the allowed modulus of the LR MI
parameter is very sensitive to the values of $x$, nevertheless the
allowed phase range of the LR  MI parameter is not changed much for
different $x$.
We find that $\mathcal{B}(B\to X_s\gamma)$  puts very strong
constraints on the upper limit of  $|(\delta_{LR}^d)_{23}|$. And
$\phi^{J/\psi}_s$ also puts very strong constraints on the lower
limit of $|(\delta_{LR}^d)_{23}|$  as well as the phase of
$(\delta_{LR}^d)_{23}$.  For $x=0.25,1,4$, the allowed spaces from
$\mathcal{B}(B\to X_s\gamma)$, $\mathcal{B}(B_{s}\to K^{-}K^{+})$
and $\Delta M_s$  are excluded by the constraint from
$\phi^{J/\psi}_s$. For $x=9$, there is small allowed space from
$\mathcal{B}(B\to X_s\gamma)$, $\mathcal{B}(B_{s}\to K^{-}K^{+})$,
$\Delta M_s$ and $\phi^{J/\psi}_s$, and it is
$|(\delta_{LR}^d)_{23}|\in[0.08,0.12]~\cup\phi_{LR}\in[20^\circ,51^\circ]$.

Previous bound $|(\delta_{LR}^d)_{23}| \leq 0.012$ for
$m_{\tilde{g}},m_{\tilde{q}}\leq600$ GeV has been  obtained from the
constraint of $\mathcal{B}(B\to X_s\gamma)$ in Ref.
\cite{Altmannshofer:2009ne}. Comparing with Ref.
\cite{Altmannshofer:2009ne},
we can see that, as shown in Fig. \ref{fig:boundsLR} (a-c),  the
bounds for the cases of $x=0.25,1,4$ from $\mathcal{B}(B\to
X_s\gamma)$ and $\mathcal{B}(B_{s}\to K^{-}K^{+})$ are stronger than
the ones only from $\mathcal{B}(B\to X_s\gamma)$ although they are
at the same order.  While, as shown in Fig. \ref{fig:boundsLR} (d)
for $x=9$ case, the constraint from $\mathcal{B}(B\to X_s\gamma)$ is
very strong and $\mathcal{B}(B_{s}\to K^{-}K^{+})$ does not give any
further constraint.

Next, we will explore  the MI SUSY effects on other observables,
which have not been (well) measured yet in $B_{s}\to
K^{(*)-}K^{(*)+}$ decays and $B^0_s-\bar{B}^0_s$ mixing, by using
the constrained parameter spaces of the LR for $x=9$ case as shown
in Fig. \ref{fig:boundsLR} (d).  The numerical results for $B_{s}\to
K^{(*)-}K^{(*)+}$ and $B^0_s-\bar{B}^0_s$ mixing are summarized in
the third column of Table \ref{Tab.SUSYpredictions}.
For $x=9$, the following comments are in order:
\begin{itemize}

\item The LR  MI can great increase
$\phi^{J/\psi}_s$ from the SM prediction range $[0.034,0.038]$ to
the SUSY prediction range $[0.16,0.52]$, which is however near to
the lower limit of the  95\% C.L. measurement.
The LR MI has been restricted by the experimental upper limit of
$\mathcal{B}(B_{s}\to K^{-}K^{+})$, and the allowed range of
$\mathcal{B}(B_{s}\to K^{-}K^{+})$ is significantly shrunken from
its SM prediction  $[9.20,45.52]\times10^{-6}$ to
$[22.80,35.30]\times10^{-6}$ by the constrained LR insertion.

\item The constrained  LR could affect the branching ratios significantly.
The allowed upper limit of $\mathcal{B}(B_s\to K^{*-}K^+)$ could be
 reduced from its SM prediction, and the allowed values of
$\mathcal{B}(B_s\to K^{-}K^{*+},K^{*-}K^{*+})$ are great increased
by the constrained  LR insertion. The range of SUSY prediction of
$\mathcal{B}(B_s\to K^{-}K^{*+})$ could differ from its SM expection
significantly.

\item The constrained LR insertion has great contributions to all mixing
CPAs in  $B_s\to K^{(*)-}K^{(*)+}$ decays, and all mixing CPAs could
be largely enhanced. In addition, the constrained LR insertion could
change $\mathcal{A}^{dir}_{CP}(B_s\&\bar{B}_s\rightarrow
 K^{*-}K^{+},K^{-}K^{*+})$ a lot.

\item  The polarization fraction $f_{L}(B_s\rightarrow
K^{*-}K^{*+})$ can be enhanced much by the constrained LR insertion.

\end{itemize}

\begin{figure}[b]
\begin{center}
\includegraphics[scale=0.89]{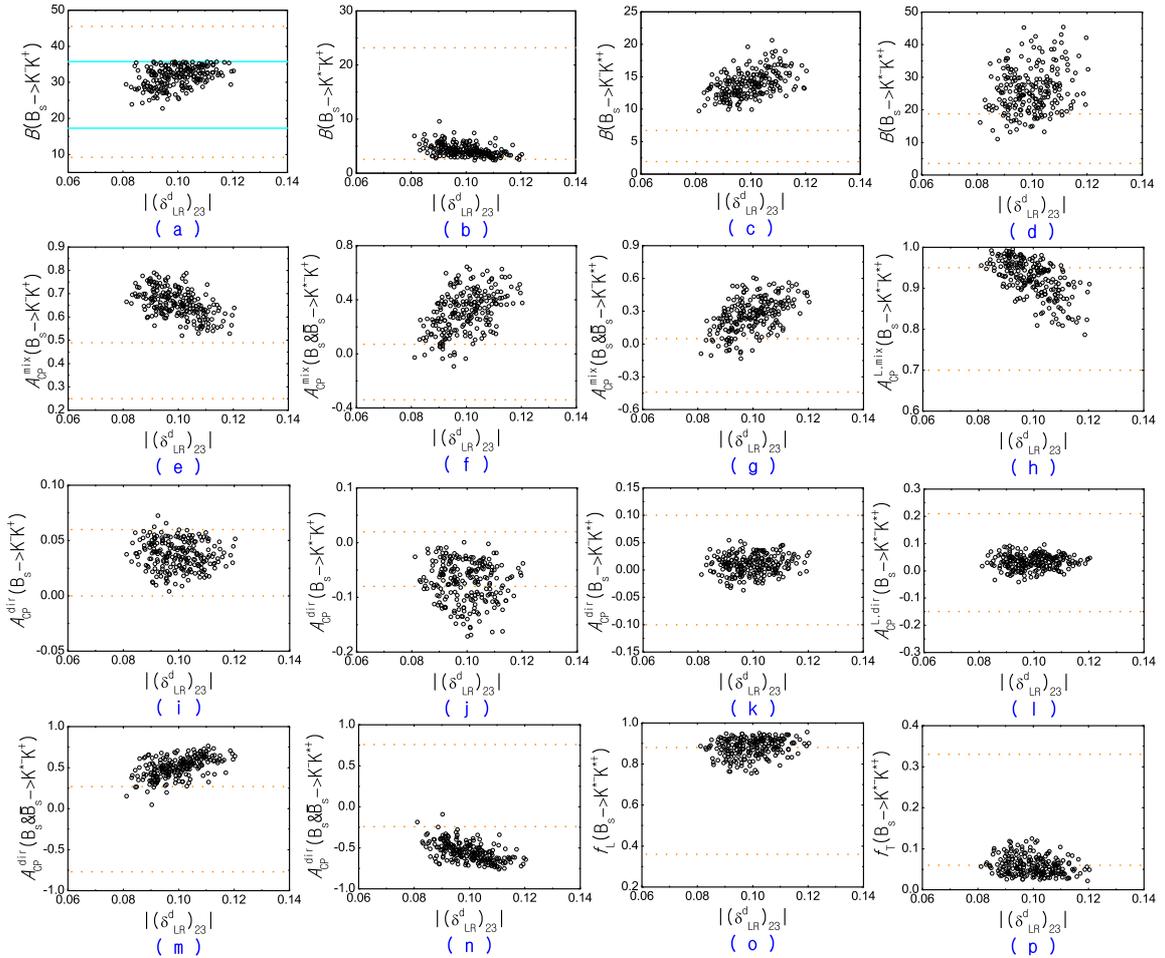}
\end{center}\vspace{-0.8cm}
\caption{\small The effects of $\left|(\delta_{LR}^d)_{23}\right|$
for $x=9$ case in $B_s\to K^{(*)-} K^{(*)+}$ decays. $\mathcal{B}$
 are in units of $10^{-6}$. The orange horizontal dash-dot lines denote the limits of SM predictions,
 and the cyan horizontal solid lines represent the $2\sigma$  error bar of the measurements.
 (The same in Fig. \ref{fig:PLR}).}
 \label{fig:MLR}
\end{figure}
\begin{figure}[ht]
\begin{center}
\includegraphics[scale=0.89]{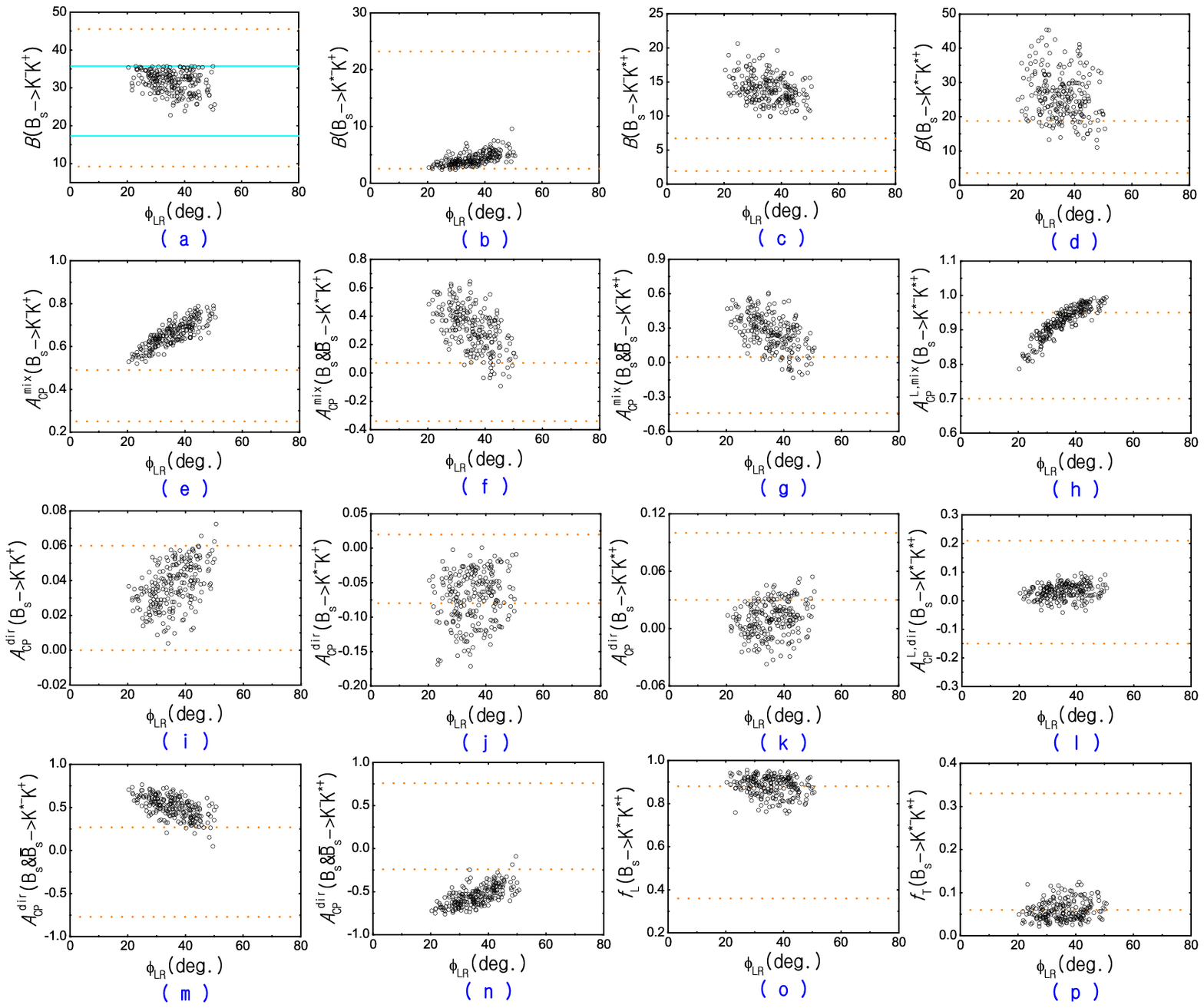}
\end{center}\vspace{-0.8cm}
\caption{\small The effects of $\phi_{LR}$ for $x=9$ case in $B_s\to
K^{(*)-}K^{(*)+}$ decays.  }
 \label{fig:PLR}
\end{figure}

 For LR insertion with $x=9$, we can present the distributions and
correlations of $\mathcal{B}$, $\mathcal{A}^{dir}_{CP}$,
$\mathcal{A}^{mix}_{CP}$,
 $f_{L,\perp}$ within the modulus or weak phase of the
constrained LR MI parameter space in Fig. \ref{fig:boundsLR} (d)  by
two-dimensional scatter plots. The LR MI effects  on all observables
of $B_s\to K^{(*)-} K^{(*)+}$ decays are displayed in Figs.
(\ref{fig:MLR}-\ref{fig:PLR}).
Fig. \ref{fig:MLR} and Fig. \ref{fig:PLR} show the sensitivities of all
observables to $|(\delta_{LR}^d)_{23}|$ and $\phi_{LR}$,
respectively. In addition, for comparing conveniently, we show the
SM bounds of these observables by orange horizontal dash lines and
the limits of the measurements
 of $\mathcal{B}(B_s\to K^-K^+)$ at 95\% C.L.  by  the cyan horizontal solid lines.
From Fig. \ref{fig:MLR} (a-d) and  Fig. \ref{fig:PLR} (a-d), one
can find that $\mathcal{B}(B_s\to K^{*-}K^+,K^{-}K^{*+})$ have mild
sensitivities to both $|(\delta_{LR}^d)_{23}|$ and $\phi_{LR}$,
while $\mathcal{B}(B_s\to K^{*-}K^{*+})$ is insensitive to
$|(\delta_{LR}^d)_{23}|$ or $\phi_{LR}$.
As shown in  Fig. \ref{fig:MLR}(e-h) and  Fig. \ref{fig:PLR}(e-h),
the LR insertion has positive effects on all four mixing CPAs, and
they are sensitive to both $|(\delta_{LR}^d)_{23}|$ and $\phi_{LR}$.
So the future measurement of any mixing CPA could further restrict
 both $|(\delta_{LR}^d)_{23}|$ and
$\phi_{LR}$.
Fig. \ref{fig:PLR} (i) and (k) show $\mathcal{A}^{dir}_{CP}(B_s\to
K^{-}K^+,K^{-}K^{*+})$ are mildly sensitive to $\phi_{LR}$. Fig.
\ref{fig:MLR} (m-n) and Fig. \ref{fig:PLR} (m-n) display that
 $\mathcal{A}^{dir}_{CP}(B_s\&\bar{B}_s\to K^{*-}K^+,K^{-}K^{*+})$
are sensitive to both $|(\delta_{LR}^d)_{23}|$ and $\phi_{LR}$.
As for the LR insertion effects on $f_L(B_s\to K^{*-}K^{*+})$ and
$f_\perp(B_s\to K^{*-}K^{*+})$, we show them in Fig. \ref{fig:MLR}
(o-p) and Fig. \ref{fig:PLR} (o-p),  we can see $f_L(B_s\to
K^{*-}K^{*+})$ and $f_\perp(B_s\to K^{*-}K^{*+})$ could be affected
significantly by the LR MI.

\subsection{RL insertion}
The SUSY contributions of the RL insertion also pick up an
$m_{\tilde{g}}/m_b$ enhancement relative to the SM. Compared to the
LR case, the RL situation is very different since the related NP
amplitude does not interfere with the SM one in $\mathcal{B}(B\to
X_s\gamma)$. The RL insertion is much more strongly constrained by
$\mathcal{B}(B\to X_s\gamma)$.

\begin{figure}[b]
\begin{center}
\includegraphics[scale=0.725]{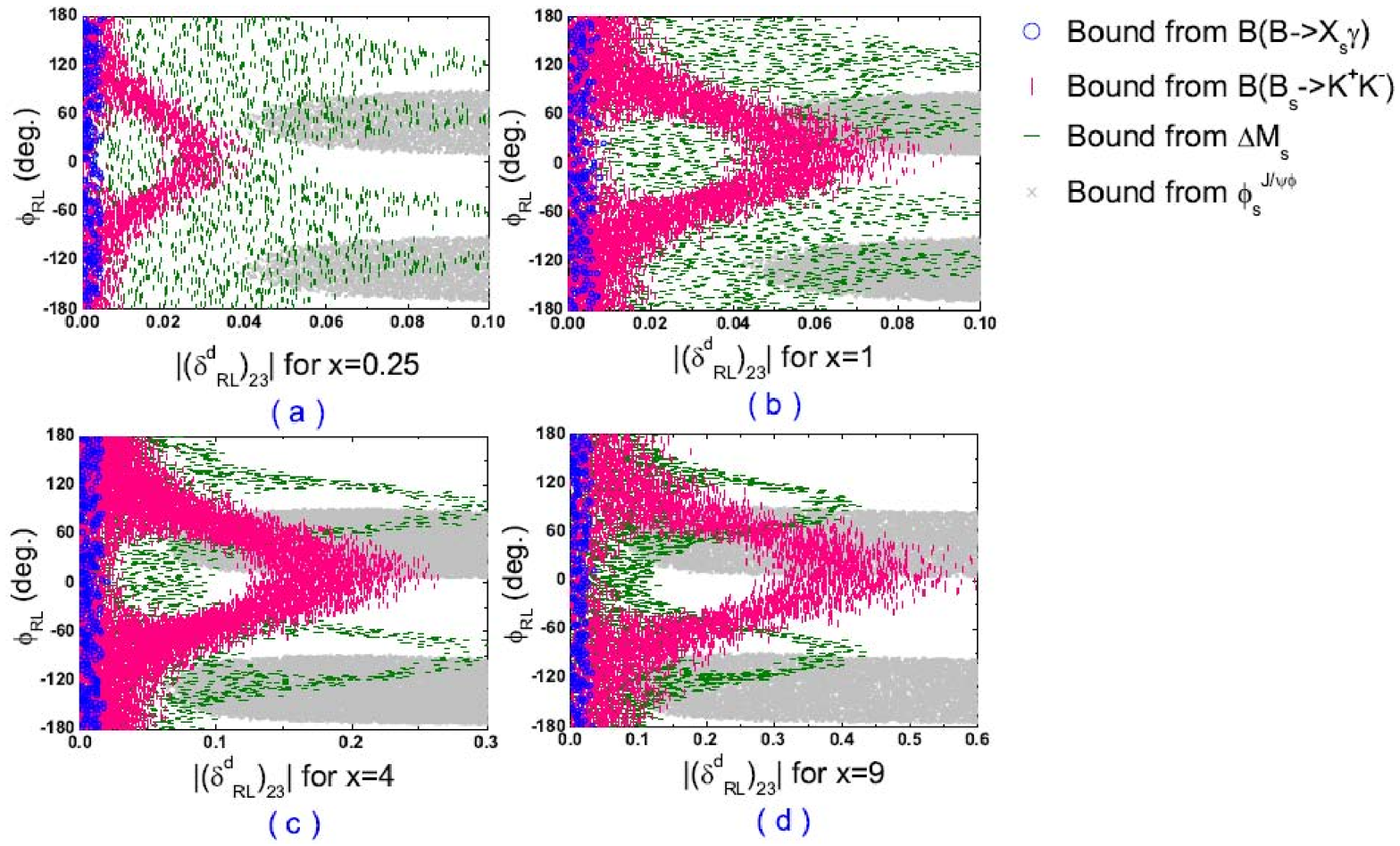}
\end{center}\vspace{-0.8cm}
\caption{ The allowed parameter spaces of the RL  MI parameters
constrained by $B_s\to K^{-}K^{+}$ decay and $B^0_s-\bar{B}^0_s$
mixing at 95\% C.L. for the squark mass $m_{\tilde{q}}=500$ GeV and
the different values of $x$.} \label{fig:boundsRL}
\end{figure}
The constrained spaces of $(\delta_{RL}^d)_{23}$ from
$\mathcal{B}(B\to X_s\gamma)$, $\mathcal{B}(B_{s}\to K^{-}K^{+})$
and $B^0_s-\bar{B}^0_s$ mixing for $m_{\tilde{q}}=500$ GeV and
different $x$ are demonstrated in Fig. \ref{fig:boundsRL}.
$\Delta\Gamma_s$ can not provide any further constraint on
$(\delta_{RL}^d)_{23}$ which is not shown in Fig.
\ref{fig:boundsRL}. As shown in this figure, there is no common
space from $\mathcal{B}(B\to X_s\gamma)$, $\mathcal{B}(B_{s}\to
K^{-}K^{+})$, $\Delta M_s$ and $\phi^{J/\psi}_s$  since
$\mathcal{B}(B\to X_s\gamma)$  puts very strong constraints on the
upper limits of $|(\delta_{RL}^d)_{23}|$, roughly
$|(\delta_{LR}^d)_{23}|\leq0.0057,0.0086,0.020,0.036$ for
$x=0.25,1,4,9$, respectively. So the RL insertion cannot accommodate
the current data of $\mathcal{B}(B\to X_s\gamma)$,
$\mathcal{B}(B_{s}\to K^{-}K^{+})$, $\Delta M_s$ and
$\phi^{J/\psi}_s$ simultaneously.

\section{Conclusions}

Motivated by the recent measurements from CDF and D{\O}
collaborations, we have studied the gluino-mediated SUSY
contributions to $B_s^0-\bar{B}_s^0$ mixing,  $B_s \to
K^{(*)-}K^{(*)+}$ and $B\to X_s\gamma$  decays with the MI
approximation. Considering the theoretical uncertainties and the
experimental error bars, we have obtained fairly constrained
parameter spaces of LL, RR, LR and RL MIs from the present
experimental data of $B_s^0-\bar{B}_s^0$ mixing, $B_s \to
K^{-}K^{+}$ and $B\to X_s\gamma$ decays. Furthermore, using the
constrained MI parameter spaces, we have predicted the MI SUSY
effects on the observables of four $B_s \to K^{(*)-}K^{(*)+}$
decays, which have not been measured yet.

For the LL and RR MIs, the strong constraint arises from
$B^0_s-\bar{B}^0_s$ mixing, and $\mathcal{B}(B_s
 \rightarrow K^-K^+)$ as well as $\mathcal{B}(B\to X_s\gamma)$ cannot
 provide any further constraint on $(\delta_{LL,RR}^d)_{23}$.
We have found that, for $x=0.25,1,4,9$ cases,  the constrained LL
and RR MIs  have little effect on the observables of $B_{s}\to
K^{(*)-}K^{(*)+}$ decays. The upper limits of $\mathcal{B}(B_{s}\to
K^{*-}K^{*+})$, $\mathcal{A}^{mix}_{CP}(B_s\&\bar{B}_s\rightarrow
K^{*-}K^{+},K^{-}K^{*+})$, $\mathcal{A}^{dir}_{CP}(B_s\rightarrow
K^{*-}K^{+})$ and $f_\perp(B_{s}\to K^{*-}K^{*+})$ are slightly
decreased from their SM values. The lower limits of
$\mathcal{A}^{L,dir}_{CP}(B_s\rightarrow K^{*-}K^{*+})$ and
$f_L(B_{s}\to K^{*-}K^{*+})$ are slightly increased from their SM
values. The allowed range of
$\mathcal{A}^{L,mix}_{CP}(B_s\rightarrow K^{*-}K^{*+})$ is enlarged.

For the LR and RL MIs, $\mathcal{B}(B\to X_s\gamma)$ puts
particularly strong constraints on the upper limits of
$|(\delta_{LR,RL}^d)_{23}|$, and $\phi^{J/\psi}_s$ also puts very
strong constraints on the lower limits of
$|(\delta_{LR,RL}^d)_{23}|$ as well as the phases of
$(\delta_{LR,RL}^d)_{23}$.
So only very narrow space of the LR MI for $x=9$ case could explain
the 95\% C.L. experimental  data of $\Delta\Gamma_s$, $\Delta M_s$,
$A^s_{SL}$, $\phi^{J/\psi\phi}_s$, $\mathcal{B}(B_s \to K^{-}K^{+})$
and $\mathcal{B}(B\to X_s\gamma)$ simultaneously. We have found the
constrained LR insertion for $x=9$ still have sizable effects on all
observables of $B_{s}\to K^{(*)-}K^{(*)+}$ decays except
$\mathcal{A}^{dir}_{CP}(B_s\to K^-K^+)$.  In addition, we have
presented the sensitivities of the observables to the constrained LR
parameter spaces in Figs. \ref{fig:MLR}-\ref{fig:PLR}. We have found
that all mixing CPAs of $B_{s}\to K^{(*)-}K^{(*)+}$ are very
sensitive to both $|(\delta_{LR}^d)_{23}|$ and $\phi_{LR}$,
moreover, $\mathcal{B}(B_s\to K^{*-}K^+,K^{-}K^{*+})$,
$\mathcal{A}^{dir}_{CP}(B_s\to K^{*-}K^+)$,
$\mathcal{A}^{dir}_{CP}(B_s\&\bar{B}_s\to K^{*-}K^+,K^{-}K^{*+})$
and $f_{L,\perp}(B_s\to K^{*-}K^{*+})$  have some sensitivities to
$(\delta_{LR}^d)_{23}|$ or $\phi_{LR}$.
So the future measurement of any mixing CPA could be very useful to
shrink/reveal/rule out the relevant LR MI parameter space. The
results could be useful for probing SUSY effects and searching
direct SUSY signals at Tevatron and LHC in the near future.

\section*{Acknowledgments}

 The work was supported by National Science Foundation
of P. R. China (Contract Nos. 11047145 and 11075059)  and Project of
Basic and Advanced Technology Research of Henan Province (Contract
No. 112300410021).

\appendix
\section*{Appendix: Input parameters}
 \label{SEC.INPUT}

The input parameters are collected in Table \ref{Tab.input}. We have
several remarks on the input parameters:
\begin{itemize}
\item \underline{Wilson coefficients}: The SM Wilson coefficients $C^{SM}_i$ are obtained from the
expressions in Ref. \cite{Buchalla:1995vs}.

\item \underline{CKM matrix element}:  For the SM predictions,
we use the CKM matrix elements from the Wolfenstein parameters of
the latest analysis within the SM in Ref. \cite{UTfit}, and   for
the SUSY predictions, we take the CKM matrix elements in terms of
the Wolfenstein parameters of the NP generalized analysis results in
Ref. \cite{UTfit}.

\item  \underline{Masses of SUSY particles}:  When we study the
SUSY effects, we will consider each possible MI
$(\delta^d_{AB})_{23}$ for $AB=LL,LR,RL,RR$ only one  at a time,
neglecting the interferences between different insertions products,
but keeping their interferences  with the SM amplitude. We fix the
common squark masses $m_{\tilde{q}}=500$ GeV and consider four
values of  $x=0.25,1,4,9$ (i.e. $m_{\tilde{g}}=250,500,1000,1500$
GeV) in all cases.
 \end{itemize}

\begin{table}[t]
\caption{Values of the theoretical input parameters. To be
conservative, we use all theoretical input parameters at 68\% C.L.
in our numerical results.} {\footnotesize
\begin{center}
\begin{tabular}{lr}\hline\hline
$m_W=80.398\pm 0.025~{\rm GeV},~m_{_{B_s}}=5.366~{\rm GeV},m_{_{K^{*\pm}}}=0.892~{\rm GeV},~m_{_{K^\pm}}=0.494~{\rm GeV},$\\
$m_t=171.3^{+2.1}_{-1.6}~{\rm
GeV},~\overline{m}_b(\overline{m}_b)=(4.20\pm0.07)~{\rm
GeV},~\overline{m}_s(2{\rm GeV})=(0.105^{+0.025}_{-0.035})~{\rm
GeV},$&\\
$\tau_{_{B_s}}=(1.472^{+0.024}_{-0.026})~{\rm ps}.$&
\cite{PDG}\\\hline
The Wolfenstein parameters for the SM predictions: &\\
$A=0.810\pm0.013,~\lambda=0.2259\pm0.0016,~\bar{\rho}=0.154\pm0.022,~\bar{\eta}=0.342\pm0.014$.&\\
The Wolfenstein parameters for the SUSY predictions: & \\
$A=0.810\pm0.013,~\lambda=0.2259\pm0.0016,~\bar{\rho}=0.177\pm0.044,~\bar{\eta}=0.360\pm0.031$.&\cite{UTfit}\\
\hline
$f_{K}=0.160~{\rm GeV},~f_{K^*}=(0.217\pm0.005)~{\rm GeV},~f^{\perp}_{K^*}=(0.156\pm0.010)~{\rm GeV},$\\
$A^{B_{s}\to  K^*}_{0}(0)=0.360\pm0.034,~A_1^{B_{s}\rightarrow
K^{\ast}}(0)=0.233\pm 0.022,
~A_2^{B_{s}\rightarrow K^{\ast}}(0)=0.181\pm 0.025,$\nonumber\\
 $ V^{B_{s}\rightarrow K^{\ast}} (0) =0.311\pm 0.026,~F^{B_{s}\to K}_{0}(0)=0.30^{+0.04}_{-0.03}.$
& \cite{BallZwicky,Duplancic:2008tk}\\\hline
$f_{B_{s}}=(0.245\pm0.025)~{\rm
GeV},~f_{B_s}\sqrt{\hat{B}_{B_s}}=0.270\pm0.030~{\rm GeV}.$ &
\cite{Lubicz:2008am}\\\hline
%
%
$\eta_{2B}=0.55\pm0.01.$&\cite{eta2B}\\\hline
$\alpha^K_1=0.2\pm0.2,~\alpha^K_2=0.1\pm0.3,~\alpha^{K^*}_1=0.06\pm0.06,~\alpha^{K^*}_2=0.1\pm0.2.$
&  \cite{Beneke:2003zv,Beneke:2006hg} \\\hline
$B_{1}^{(s)}(m_b)=0.86(2)\left(^{+5}_{-4}\right),~~B_{2}^{(s)}(m_b)=0.83(2)(4),~~B_{3}^{(s)}(m_b)=1.03(4)(9)$,&\\
$B_{4}^{(s)}(m_b)=1.17(2)\left(^{+5}_{-7}\right),~~B_{5}^{(s)}(m_b)=1.94(3)\left(^{+23}_{-7}\right)$.&\cite{Bparameter}\\\hline
\end{tabular}
\end{center}}\label{Tab.input}
\end{table}

\end{document}